# Educational assortative mating and household income inequality: evidence from Brazil, Indonesia, Mexico, and South Africa

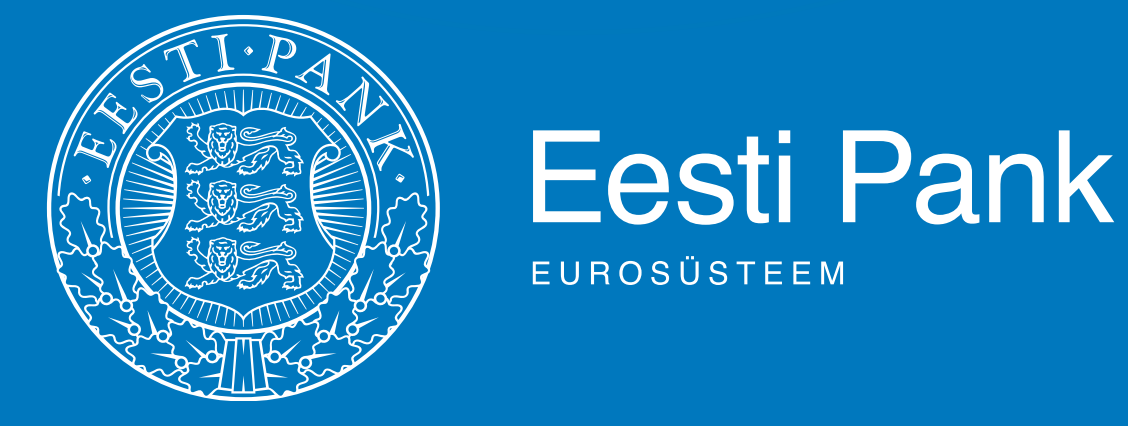

# Educational Assortative Mating and Household Income Inequality: Evidence from Brazil, Indonesia, Mexico, and South Africa

Ana Kujundzic*

**Abstract**

This paper presents new empirical evidence from four emerging economies on the relationship between educational assortative mating and household income inequality. It also introduces a new methodological approach, first proposed in my earlier work, that overcomes limitations of standard sorting measures used in the literature and allows for the identification of pure sorting trends without imposing restrictive assumptions about the sorting process. I find that people in Brazil, Indonesia, Mexico, and South Africa tend to sort into internally homogeneous marriages based on education level. While educational sorting has a noticeable impact on household income inequality in any given year, changes in the degree of sorting over time barely have any impact on inequality. Further analysis reveals that this counterintuitive result is due to different dynamics within educational groups. The inequality-decreasing impact from reduced sorting among the highly educated is almost entirely offset by the inequality-increasing impact from increased sorting among the least educated. While it is certainly reassuring that concerns about educational assortative mating having a potentially large effect on income disparities between households appear to be unwarranted, these findings suggest another concerning narrative. Marginalization processes are occurring at low levels of educational distribution. The least educated are being left behind, facing limited labor market opportunities and diminished chances of achieving upward socio-economic mobility through marriage to more educated partners.

---

* Ana Kujundzic was employed as a visiting researcher of Eesti Pank. Corresponding author's email: akujundzic.1@gmail.com.

The author is thankful to Peter Lanjouw, Erwin Bulte, Janneke Pieters, Rein Haagsma, Jaanika Meriküll, and Merike Kukk for their valuable feedback and suggestions for improving this paper. Ana Kujundzic also greatly appreciates the comments from the participants at the Eleventh ECINEQ Conference, and seminar participants at Eesti Pank, London School of Economics and Political Science, and Wageningen University.

## Non-technical summary

Income inequality has become a central global concern. Despite decades of economic growth and rising average living standards, large income gaps have persisted and, in many cases, widened. Excessive levels of inequality are problematic because they can weaken social cohesion, fuel political polarization, slow poverty reduction, and ultimately hinder long-term economic growth. Understanding what drives growing inequalities and what can be done about it is therefore a key policy challenge.

This paper examines one potential but often overlooked contributor to income inequality: the question of who marries whom, a choice that sits at the core of household formation and has long-term implications for income and wealth accumulation. In particular, the paper focuses on *educational assortative mating*, the tendency for people to marry partners with similar levels of education. Because education strongly influences earnings and employment opportunities, marriages between two highly educated individuals can concentrate income within certain households, while marriages between two low-educated individuals may reinforce economic disadvantages. This raises a natural concern: could educational sorting be an important driver of income inequality between households?

The study addresses this question in the context of four large middle-income countries – Brazil, Indonesia, Mexico, and South Africa – which are especially relevant settings for this analysis because they have persistently high levels of income inequality and have experienced rapid educational expansion over recent decades. They also account for a substantial share of the world's population living in extreme poverty. Yet, compared with high-income countries, they have received relatively little attention in research on marital sorting and inequality.

The paper asks three main questions. First, do individuals in these countries tend to marry partners with the same education level? Second, has this tendency changed over time? Third, to what extent does educational assortative mating affect income inequality between households?

A key contribution of the paper is methodological. Much of the existing literature struggles to properly measure educational sorting trends. When education expands rapidly – as it has in many countries since the 1970s – standard measures can misleadingly suggest that sorting has increased even if underlying sorting patterns have not changed. This paper applies a new method, first proposed in my earlier work, that explicitly controls for changes in educational attainment, allowing for the identification of changes in pure sorting.

I find that although individuals in all four countries tend to marry partners with the same education level, this tendency has decreased over time. Moreover, while educational sorting has a noticeable impact on household income inequality in any given year, changes in the degree of sorting over time barely have any impact on inequality. The reason for this has to do with different group dynamics. Sorting among the highly educated has weakened, which would tend to reduce inequality, while sorting among the least educated has strengthened, which pushes inequality in the opposite direction. These offsetting forces largely cancel each other out at the aggregate level.

These results offer a reassuring message and a warning at the same time. They suggest that fears of educational assortative mating dramatically worsening household income inequality may be overstated. At the same time, they point to a more troubling pattern: increasing social and economic isolation among the least educated, who face limited labor market opportunities and shrinking prospects for upward mobility through marriage. Addressing inequality, therefore, may require greater attention to marginalization processes that are happening at the bottom of the educational distribution.

## 1. Introduction

High levels of income inequality – both within and between countries – have become such a pressing concern that reducing it was included for the first time as a key component of Goal 10 in the 2030 Agenda for Sustainable Development, adopted by the United Nations (UN) Member States in 2015. In the 2020 edition of their World Social Report, the UN reports that "the extraordinary economic growth and widespread improvements in well-being observed over the last several decades have failed to close the deep divides within and across countries." It goes on to note how income inequality has increased in most developed countries and in some middle-income countries, including China and India, since 1990 – and that countries where inequality has grown are home to more than two-thirds of the world population. While some inequality is inevitable in a market-based economic system due to differences in talent, effort, and luck, excessive levels of inequality are problematic because they can erode social cohesion, increase political polarization, hinder poverty reduction, and ultimately slow economic growth (Rodrik 1999; Berg et al. 2018).

An important question is what is fueling growing inequalities and what can be done about it. Some of the drivers that have been identified and extensively debated in the literature include globalization, technological change, and educational meritocracy. For example, Goldberg and Pavcnik (2007) and Bourguignon (2015) argue that while globalization promotes economic growth and poverty reduction, it also exacerbates income inequality within countries due to its uneven distribution of benefits – disproportionally favoring the wealthy over the poor. Acemoglu and his co-authors contend that skill-biased technological change has led to job polarization by automating routine tasks traditionally performed by middle-skill workers, thereby increasing income inequality (Acemoglu and Autor 2011; Acemoglu and Restrepo 2022, 2024; Acemoglu and Loebbing 2024). Finally, Sandel (2020) critiques the concept of educational meritocracy, arguing that it perpetuates existing inequalities by providing unfair advantages to wealthy families who can afford superior education and resources for their children.

This paper explores another potential driver of income inequality: the prevalence of homogamy, also known as positive assortative mating, where individuals of similar socioeconomic status partner with each other, thereby widening the income gap between households. I specifically focus on educational sorting because education is considered an important acquired trait on which people sort, given its role as a leading determinant of employment and earnings. That said, this study addresses the following questions in the context of four middle-income countries: Do individuals tend to sort into internally homogenous marriages based on education level? If so, how has assortative mating evolved over time in these countries? What is the impact of educational assortative mating on household income inequality?

The importance of marital sorting for understanding key economic outcomes, such as between-household income inequality, was almost completely ignored by economists until Gary Becker's seminal 1973 paper on the economics of marriage and family. Becker's work is the first systematic analysis of marriage from an economic perspective, laying the foundation for research in family economics. One of the main concepts introduced by Becker is that the question of who marries whom can be analyzed in terms of the complementarity and substitutability of inputs used in the production of household commodities. Individuals behave as if they maximize the amount of household-produced commodities they would receive if they marry, subject to the full income constraint. If individual traits of men and women are treated as inputs into the household production function, positive assortative mating, or the association of likes, is an optimal solution when traits are complements. Conversely, negative assortative mating, or the association of unlikes, is optimal when traits are substitutes. Becker predicts that marital patterns will mostly exhibit positive sorting since traits are typically, but not always, complements in the sense that they have a reinforcing effect on household productivity (e.g., intelligence, age, education, family background).

In his follow-up to the 1973 paper, Becker discusses the implications of positive assortative mating for household income inequality, concluding that sorting on acquired traits such as education can contribute to widening income disparities between families. This is because education is a key determinant of earning potential and economic resources, with highly educated individuals likely to earn significantly more than those with lower levels of education (Becker 1974).

Most empirical research on educational sorting has been conducted in the context of the United States (US) and Western European countries, with findings consistent with Becker's prediction of positive assortative mating by education level (e.g., Pencavel 1998; Schwartz and Mare 2005; Blossfeld 2009; Greenwood et al. 2014; Chiappori et al. 2017; De Hauw et al. 2017; Eika et al. 2019). Evidence regarding the impact of educational positive assortative mating (hereafter educational PAM) on household income inequality is more scarce. Most studies in this area find that while educational PAM can affect household income inequality, changes in educational PAM over time have had little effect on changes in income inequality between households (e.g., Breen and Salazar 2011; Greenwood et al. 2015; Hryshko et al. 2017; Boertien and Permanyer 2019; Eika et al. 2019).

This paper contributes to the existing empirical literature in two ways. First, it investigates whether the findings from the US and Western European countries generalize to four major emerging economies: Brazil, Indonesia, Mexico, and South Africa. To my knowledge, only a handful of studies have examined the relationship between educational sorting and income inequality in some of these countries, but not all four simultaneously (e.g., Fernandez et al. 2005; Samudra and Wisana 2015; Hoehn-Velasco and Penglase 2023). Additionally, these studies use different methodologies and datasets from this one.

Emerging economies tend to have higher levels of income inequality than high-income countries and have experienced rapid educational expansion over the past couple of decades. Using OECD-comparable estimates of equivalized household disposable income, Balestra et al. (2018) show that Gini coefficients in major emerging economies such as Brazil, Indonesia, and South Africa are close to 0.50 or even above 0.60, whereas the average among OECD member countries is around 0.32 and even the five most unequal OECD countries are around 0.40. Income inequality in emerging economies has remained at these high levels since the early 1990s. While Mexico is an OECD member country, it remains one of the most unequal countries in Latin America – the most unequal region in the world – with income Gini coefficients typically between 0.45 and 0.50, well above the OECD average. As emphasized in the UN World Social Report (2020), such persistently high levels of income inequality are rooted in structural characteristics of these economies, including deep regional divides, widespread labor market informality, and large educational and ethnic disparities.

These four countries also stand out for having undergone substantial educational expansion, with large declines in low-education groups and rapid growth in secondary and tertiary attainment (Ritchie et al. 2023). This expansion is clearly reflected in Table 3 in Section 3.2 and in the changing marginal distributions of education levels in the contingency tables presented in Tables A.1–A.4 in Appendix A. For example, in Brazil, about 98% of married women had less than secondary education in 1970, compared with 63% in 2010, while the share with secondary or tertiary education increased from 2 to 37%. Indonesia shows an even more dramatic shift. The share of married women in the two lowest educational categories fell from 74 to 35% between 1993 and 2014, and senior-secondary or tertiary attainment grew from 15 to 43%. Mexico experienced a similarly large expansion, with the share of married women with less than secondary education declining from 99% in 1970 to 77% in 2015, while secondary or tertiary attainment rose from 1 to 23%. In South Africa, the share of married women in the two lowest educational groups fell from 73% in 1996 to 51% in 2011, while secondary and tertiary attainment nearly doubled, increasing from 27 to 49%. Changes in educational attainment of married men follow the same broad pattern in each country, reflecting widespread educational expansion for both men and women.

This combination of unusually high income inequality and rapid educational expansion makes these countries particularly relevant settings for examining the potential impact of educational sorting on household income inequality. Moreover, according to the UN World Social Report (2020), these four countries together are home to roughly 40–50 million people living in extreme poverty, representing about 6–7% of the 727 million people worldwide who lived on less than $1.90 a day in 2015. This underscores the importance of studying income inequality in these countries from a poverty-reduction perspective.

The second contribution of this paper is methodological. I illustrate a new method for measuring trends in assortative mating, first proposed in Kujundzic (2024). As discussed in more detail in my 2024 paper, there is no consensus in the existing literature on how to measure differences in educational sorting across time and space. This lack of consensus arises largely because most measures of sorting used in this literature fail to distinguish between changes (or differences) in educational attainment (i.e., educational expansion) and changes (or differences) in pure sorting. When a sorting measure cannot distinguish between these two components, researchers might incorrectly conclude that the degree of educational sorting has increased over time and that it has significantly contributed to household income inequality, when in fact educational expansion – rather than sorting – may be associated with rising inequality, all else being equal. My new proposed method, illustrated in this paper, addresses this limitation by controlling for changes in educational attainment, thereby allowing for the identification of pure sorting trends.

I find evidence of PAM for every educational group in each year and country examined: Brazil (1970 and 2010), Indonesia (1993 and 2014), Mexico (1970 and 2015), and South Africa (1996 and 2011) (see Tables 5–8 and Figure 1 in Section 4.1) These findings are consistent with Becker's theory of sorting and with findings in most empirical studies of educational assortative mating in both high-income and low- and middle-income countries. Regarding group dynamics, I find that highly educated individuals sort among themselves the most, while the least educated sort the least in every country-year combination. Although the level of sorting among the highly educated is the highest, it has been decreasing over time in all four countries. At the same time, the least educated have been increasingly sorting into internally homogenous marriages in all countries except Brazil (see Figure 4 in Section 4.4).[1] Interestingly, Eika et al. (2019) reach the same conclusion about group sorting trends for a set of high-income countries: Denmark (1980–2013), Germany (1984–2013), Norway (1967–2013), the United Kingdom (1979–2013), and the US (1940–2013).

After controlling for changes in educational attainment using the method proposed in Kujundzic (2024), I find that overall PAM across all education levels has decreased in all four countries during the study period (see Figure 2 in Section 4.1).[2] The magnitude of the decrease varies across countries, ranging from 14% in Indonesia and South Africa to 23% in Mexico and

[1] I reach this conclusion after controlling for changes in educational attainment of men and women using the method proposed in Kujundzic (2024).

[2] Overall PAM is defined as a weighted sum of the PAM level for each educational group.

35% in Brazil. A recent study by Boertien and Permanyer (2019) also finds that overall PAM has decreased in most of the 20 European countries they examined, covering the period from the 1980s to the 2010s, and in the US from 1974 to 2016. However, Eika et al. (2019) find that overall PAM steadily increased in the US from 1940 to the mid-1980s, after which it changed relatively little. Some studies in the context of the US (e.g., Rosenfeld 2008) argue that overall educational PAM has remained relatively stable over time, while others, like Schwartz and Mare (2005), claim it has increased. These differing findings illustrate the ongoing lack of consensus in the literature regarding educational sorting trends.

I closely follow the method outlined in Greenwood et al. (2014) to test the hypothesis that educational PAM can exacerbate income inequality between households. Specifically, I consider two hypothetical scenarios: a complete absence of sorting ("random matching" scenario) and a different degree of sorting than that observed in data for a particular year. In the first simulation, I find that household income inequality would be lower than observed if people matched randomly instead of assortatively. This is true for all four countries and every year examined. In the second simulation, I examine what the level of household income inequality in period 2 would be if the degree of sorting were the same as in period 1. Given the hypothesis of a positive correlation between educational sorting and household income inequality, and the findings that overall PAM has decreased in all four countries, we would expect inequality to be higher in period 2 if individuals sorted as they did in period 1. Contrary to expectations, I find that the simulated level of inequality in period 2 barely differs from the actual levels observed in period 2, with differences of less than 3%. These findings are in line with conclusions from Greenwood et al. (2015) for the US, Eika et al. (2019) for the US and Norway, and Boertien and Permanyer (2019) for the US and 20 European countries.

Additional analysis reveals that this counterintuitive result is due to differing group dynamics, with sorting trends by educational group varying in both magnitude and direction. This is perhaps best illustrated in the case of South Africa, where the inequality-decreasing impact of a decrease in sorting among the highly educated is almost completely offset by the inequality-increasing impact of an increase in sorting among the least educated group. As a result, a 14% decrease in overall PAM leads to only 1% reduction in household income inequality.

The rest of the paper is organized as follows. Section 2 outlines the measures of assortativeness and household income inequality, and details two simulation procedures used to quantify the impact of sorting on household income inequality. Section 3 discusses the data sources, sample restrictions, and variable definitions. The main findings and their comparisons with other empirical studies are presented in Section 4. The last section summarizes the main findings and offers concluding remarks.

## 2. Methodology

The empirical methodology for measuring the degree of educational PAM and its effect on household income inequality relies on the use of contingency tables to describe and analyze observed mating patterns. A contingency table is a matrix that displays frequency counts or relative frequencies of occurrences for each unique combination of the row and column categorical variables. This section begins by describing how a contingency table is used within the context of the educational marriage market to define and measure the degree of educational PAM. It then explains the calculation of observed and counterfactual Gini coefficients as a function of different household types situated in different percentiles of the household income distribution. While I closely follow the approach in Greenwood et al. (2014) to compute observed and counterfactual Gini coefficients in Sections 2.2 and 2.3, I apply my new Quadratic Programming (QP) method to standardize contingency tables. As explained in Section 2.3.2 and in more detail in Kujundzic (2024), table standardization is of utmost importance for comparative studies of educational assortative mating because it isolates changes in pure sorting from changes in the educational distribution of men and women.

## 2.1 Measure of educational PAM

I measure the extent of educational assortative mating by examining an $(I \times J)$ contingency table, where each cell gives the observed share (or relative frequency) of household type $t_{ij}$ with husbands with education level $i \in \{1, \ldots, I\}$ and wives with education level $j \in \{1, \ldots, J\}$ (denoted by $f_{t_{ij}}$).[3] The row-specific total of the contingency table (denoted by $f_{t_{i+}}$) is the share of household types with husbands with education level $i$. The column-specific total (denoted by $f_{t_{+j}}$) is the share of household types with wives with education level $j$. The last column (row) represents the marginal distribution of education levels of husbands (wives).[4] The sum of all cells equals 1, representing 100% of households in the sample. Additionally, the sum of row-specific totals equals the sum of column-specific totals, both of which must equal 1.

The measure of assortativeness is the interest factor (I), also known as the likelihood ratio, which is defined as the probability of observing a household type $t_{ij}$ compared to the probability of observing the same type if matching were random:

$$I_{ij} = \frac{P(t_{ij})}{P(t_{i+})P(t_{+j})} = \frac{f_{t_{ij}}}{f_{t_{i+}} f_{t_{+j}}} \quad (1)$$

The probability of observing a household type $t_{ij}$ is represented by the $(i, j)$ cell of the relative contingency table ($f_{t_{ij}}$). Random matching implies statistical independence between the educational levels of spouses, meaning that the probability of observing a household type $t_{ij}$ if matching were random is equal to the product of the $i$ and $j$ group marginals ($f_{t_{i+}} f_{t_{+j}}$). If individuals with the same education level marry more frequently compared to the benchmark outcome of random matching, the values of $I_{ij}$ on the main diagonal of the contingency table would be greater than 1. Thus, the measure of PAM for each educational group is the positive deviation of $I_{ij}$ from unity in each cell on the main diagonal.

The measure of overall PAM across all educational groups is a weighted sum of the values of $I_{ij}$ along the main diagonal, with the weights corresponding to the marginal distribution of wives' education. It is worth noting that one could alternatively use the marginal distribution of husbands' education as weights, or any convex combination of the marginal distributions of husbands and wives.

[3] Each inner cell of the relative contingency table can be interpreted as the probability of observing a household type $t_{ij}$.

[4] The row and column marginals (i.e. row and column totals) of a relative contingency table can be interpreted as the probabilities of observing households with husbands with education level $i$ and wives with education level $j$, respectively.

### 2.2 Measure of household income inequality

The metric used to measure household income inequality is the Gini coefficient, calculated from the Lorenz curve of household income distribution. Let $f_{pt_{ij}}$ denote the share of household type $t_{ij}$ in income percentile $p$, and $y_{pt_{ij}}$ denote the share of total income earned by type $t_{ij}$ in income percentile $p$. The Lorenz curve is derived by plotting the cumulative share of all household types at percentile $p = k$ on the x-axis:

$$F_k = \sum_{p=1}^{k} \sum_{t_{ij}} f_{pt_{ij}} \tag{2}$$

against the cumulative share of total income earned by all household types at percentile $p = k$ on the y-axis:

$$Y_k = \sum_{p=1}^{k} \sum_{t_{ij}} y_{pt_{ij}} \tag{3}$$

For $n$ percentiles, the Gini coefficient[5] equals:

$$G = \sum_{k=1}^{n} \left[ F_k Y_{(k+1)} - F_{(k+1)} Y_k \right] \tag{4}$$

I work with $n = 10$ percentiles (deciles) to make the counterfactual analysis more tractable. Although working with deciles reduces the variability in inequality, resulting in a slightly lower Gini coefficient than if $n = 100$, the exact value of the Gini coefficient is arbitrary for the purpose of the counterfactual analysis.

### 2.3 Counterfactual analysis

To test the hypothesis that educational PAM can exacerbate income inequality between households, I compare the observed level of household income inequality with the levels that would occur under two counterfactual scenarios. The first scenario is the extreme case of a complete absence of sorting, where people match randomly. In this scenario, I ask what household income inequality would look like in each year if people matched randomly instead of assortatively. The second scenario is the case of PAM with a different degree of sorting than that observed in the data. Here, I ask what household income inequality would look like in the second period if people matched as in the first period. In both counterfactual simulations, the household income distribution is taken as given and only the joint distribution of spouses' education is altered, isolating a partial-equilibrium effect of educational PAM on household income inequality. This approach abstracts from potential behavioral responses – such as labor supply adjustments within couples – so the estimates should be interpreted as capturing the compositional effect of sorting, all else being equal.

[5] In their online appendix, Greenwood et al. (2014) derive this expression for the Gini coefficient from the conventional formula $G = 1 - 2\Delta$, where $\Delta$ represents the area below the Lorenz curve.

### 2.3.1 Case 1: random matching

To calculate the counterfactual Gini coefficient for the random matching scenario, the observed share of household type $t_{ij}$ in income percentile $p$ ($f_{pt_{ij}}$) in Equation (2) must be replaced with the share that would occur under random matching (denoted by $\tilde{f}_{pt_{ij}}$). This replacement involves rescaling $f_{pt_{ij}}$ by the scaling factor $1/I_{ij}$, as suggested by Greenwood et al. (2014):

$$\tilde{f}_{pt_{ij}} = \left(\frac{1}{I_{ij}}\right) f_{pt_{ij}} = \left(\frac{f_{t_{i+}} f_{t_{+j}}}{f_{t_{ij}}}\right) f_{pt_{ij}} \tag{5}$$

Similarly, the observed share of total income earned by household type $t_{ij}$ in income percentile $p$ ($y_{pt_{ij}}$) in Equation (3) is also rescaled by the scaling factor $1/I_{ij}$ to obtain the counterfactual household income distribution under the random matching scenario:

$$\tilde{y}_{pt_{ij}} = \left(\frac{1}{I_{ij}}\right) y_{pt_{ij}} = \left(\frac{f_{t_{i+}} f_{t_{+j}}}{f_{t_{ij}}}\right) y_{pt_{ij}} \tag{6}$$

The counterfactual Gini coefficient is then computed by plugging the cumulative counterfactual distribution of household types ($\tilde{F}_k$) and their incomes ($\tilde{Y}_k$) into Equation (4).

To better understand how the rescaling procedure works, let us examine the $I_{ij}$ table for the 2010 Brazil sample in Panel B of Table 5 in Section 4.1. The $I_{ij}$ value for household type $t_{11}$, where both spouses have less than a primary school education, is $I_{11} = 1.86$. The scaling factor for type $t_{11}$ households across all income percentiles is $1/I_{11} = 0.54$. This means that, under random matching, the observed share of type $t_{11}$ households is reduced by a factor of 0.54 in every income percentile. Similarly, the observed share of total income earned by type $t_{11}$ households is reduced by a factor of 0.54 in every income percentile. Each household type has its own specific scaling factor. For the 2010 Brazil sample, there are 16 scaling factors for 16 possible household types. By repeating the scaling operation for the other 15 types, we obtain the counterfactual distribution of household types and their incomes, which is then used to calculate the counterfactual Gini coefficient

### 2.3.2 Case 2: Change in the degree of sorting

In the second simulation, I calculate the counterfactual Gini coefficient for the second time period ($t_2$) while keeping the degree of sorting as it was in the first period ($t_1$). In other words, I ask what household income inequality would look like in the second period if people matched as in the first period. Since identifying the true degree of sorting in the first period requires controlling for differences in educational attainment between the two periods, I first standardize the first-period contingency table so that it has the same marginals as the second-period table. I then compute the counterfactual Gini coefficient using the standardized first-period table.

As I discuss in more detail in Kujundzic (2024), contingency table standardization allows for identification of pure sorting trends by separating them from mechanical changes in the marginals (i.e., educational expansion in the context of this paper) and is therefore crucial not only for this simulation exercise but also for the analysis of sorting trends. Table standardization controls for changes in the marginals by transforming the initial table so that it has the desired (target) marginals while, importantly, preserving the core pattern of association of the initial table during the standardization process. Another way to think about table standardization is as a decomposition procedure in which the observed difference in the overall level of sorting is decomposed into differences in pure sorting and differences in the marginals. After standardization, the tables have identical (uniform) marginals, making the marginal differences component equal to zero. Consequently, any observed difference in overall sorting can be attributed solely to differences in pure sorting.

Given that I use the interest factor (I) as a measure of educational sorting, I standardize the first-period table using the new Quadratic Programming (QP) method proposed in Kujundzic (2024). This new standardization method adjusts the marginals of the first-period table so that they match the marginals of the second-period table, while preserving the core pattern of association of the initial first-period table as measured by local interest factors ($I_{ij}$). The QP standardization method is explained in Appendix B, and the standardized first-period tables are reported in Panel C of Tables A.1–A.4 in Appendix A.

I use these standardized first-period tables in my calculation of the counterfactual Gini coefficients. The simulation procedure is the same as in the case of random matching, except that now the observed share of household type $t_{ij}$ in income percentile $p$ in the second period ($f_{pt_{ij}}^{t_2}$) is replaced with the share observed in the first period ($f_{pt_{ij}}^{t_1}$). This is achieved with the following rescaling operation:

$$\hat{f}_{pt_{ij}}^{t_2} = \left( \frac{f_{t_{ij}}^{t_1}}{f_{t_{ij}}^{t_2}} \right) f_{pt_{ij}}^{t_2} \tag{7}$$

with the scaling factor being the ratio of the $(i, j)$ cells of the standardized first-period table and the observed second-period table.

Likewise, the observed share of total income earned by household type $t_{ij}$ in income percentile $p$ in the second period ($y_{pt_{ij}}^{t_2}$) is rescaled using the same scaling factor to obtain the counterfactual household income distribution:

$$\hat{y}_{pt_{ij}}^{t_2} = \left( \frac{f_{t_{ij}}^{t_1}}{f_{t_{ij}}^{t_2}} \right) y_{pt_{ij}}^{t_2} \tag{8}$$

The counterfactual Gini coefficient is computed by plugging the cumulative counterfactual distribution of household types ($\hat{F}_k$) and their incomes ($\hat{Y}_k$) into Equation (4).

To illustrate this procedure, consider again the case of Brazil. The standardized 1970 contingency table is used as the first-period table and the observed 2010 contingency table is used as the second-period table (see panels C and B of Table A.1 in Appendix A). The scaling factor for type $t_{11}$ households is $0.338/0.261 = 1.3$. This means that the observed share of $t_{11}$ types in 2010 is increased by a factor of 1.3 in every income percentile. Likewise, the observed share of total income earned by $t_{11}$ types is increased by the same factor in every income percentile. As before, each household type has its own specific scaling factor. In the case of Brazil, there are 16 scaling factors for 16 possible household types. Repeating the scaling operation for the other 15 types produces the counterfactual distribution of household types and their incomes, which is then used to calculate the counterfactual Gini coefficient

## 3. Data

In what follows, I describe data sources, sample restrictions, and the main variables used in the analysis.

### 3.1 Data sources

Data for Indonesia is from the first and last wave of the Indonesian Family Life Survey (IFLS), fielded in 1993 and 2014. Given the size and terrain of Indonesia, along with survey cost considerations, the baseline sample was selected to maximize representation and capture the cultural and socioeconomic diversity of the country, resulting in 83% representation of the Indonesian population (Strauss and Witoelar 2021).

Data for Brazil, Mexico, and South Africa are random (representative) samples from population censuses provided by the national statistical offices of the participating countries to the International Integrated Public Use Microdata Series (IPUMS-International).[6] A sample size of 10%, for example, is a 10% random extract from the population census. IPUMS-International harmonizes this microcensus data to enable cross-national and cross-temporal comparisons. Census years and sample sizes used in this paper are shown in Table 1. Given the availability of data for the main variables used in the analysis, I use the longest possible time span to study long-run educational sorting trends.

**Table 1. Census years and sample sizes for the IPUMS-International countries**

| Country | Census Year | Sample size (% extract from population census) |
|---|---|---|
| Brazil | 1970 | 25.0 |
| Brazil | 2010 | 10.0 |
| Mexico | 1970 | 1.0 |
| Mexico | 2015 | 9.5 |
| South Africa | 1996 | 10.0 |
| South Africa | 2011 | 8.6 |

*Notes:* The earliest census year available in the IPUMS-International for South Africa is 1996. IPUMS-International census data are random samples of population censuses provided by participating countries. A sample size of 10%, for example, is a 10% random extract from the population census.

### 3.2 Sample restrictions and variables definition

Considering the type of data used in this paper, it is important to first address some concerns that arise in empirical studies of sorting patterns when information on the number and timing of marriages is unavailable. Most census data include information only on currently intact marriages at the time of the census, encompassing both first marriages and remarriages formed earlier. Analyses limited to current marriages may lead to biases if divorce rates are higher among heterogamous couples, resulting in a disproportionally higher share of homogamous couples among intact marriages in the census data (Zhang and van Hook 2009).

[6] The author wishes to acknowledge the national statistical offices of Brazil, Mexico, and South Africa for providing the original data to the IPUMS-International.

That said, divorce rates are historically much lower in Brazil, Mexico, and South Africa compared to the US and European countries (Herre et al. 2020). The average divorce rate for working age individuals (ages 25–55) in all three countries in IPUMS-International micro-census samples is about 5%, which is small enough to alleviate concerns about the potential overestimation of the number of homogamous marriages due to census data limitations (see Table 2).

**Table 2. Marital status of working age individuals**

| **Panel A: Brazil (ages 25–55)** | 1970 | 2010 |
|---|---|---|
| Single / never married | 14.95% | 15.24% |
| Married | 78.05% | 70.54% |
| Divorced / separated | 3.25% | 12.63% |
| Widowed | 3.63% | 1.59% |
| **Panel B: Indonesia (ages 15-55)** | 1993 | 2014 |
| Single / never married | 32.25% | 22.28% |
| Married | 62.60% | 72.11% |
| Divorced / separated | 2.38% | 2.77% |
| Widowed | 2.76% | 1.73% |
| **Panel C: Mexico (ages 25–55)** | 1970 | 2015 |
| Single / never married | 12.01% | 14.93% |
| Married | 81.91% | 76.21% |
| Divorced / separated | 2.56% | 6.60% |
| Widowed | 3.51% | 1.89% |
| **Panel D: South Africa (ages 25–55)** | 1996 | 2011 |
| Single / never married | 32.00% | 41.06% |
| Married (monogamous) | 58.70% | 53.36% |
| Divorced / separated | 4.00% | 2.97% |
| Widowed | 2.68% | 2.61% |

Unlike the IPUMS-International census data, the IFLS data includes information on the number and timing of marriages, and therefore, only couples in their first marriages are included in the Indonesian sample to reduce biases associated with marital dissolution.

Brazil, Mexico, and South Africa samples are restricted to working age married individuals between 25 and 55 years old. This age restriction is common in the literature because most individuals in this age group have completed their education and are not relying on government pensions or similar benefits (e.g., Hyslop and Mare 2005; Daly and Valletta 2006; Chiappori et al. 2017). The lower age limit for the Indonesian sample is 15 because the IFLS provides information on employment and wages for adults 15 years and older who are out of school.

A household is defined as a married monogamous couple with or without children living together at the time of the interview. Households with more than one married couple and single-headed households are excluded from the analysis. Additionally, households with missing information for one or both spouses on age, educational attainment, and income are also excluded.[7]

[7] Households with zero income values are included in the analysis.

Individuals are grouped into different education levels based on years of schooling completed and degrees obtained. In the case of Indonesia, I construct the education level variable that reflects the five main stages of Indonesia's educational system: less than primary school (less than six years of schooling completed at the time of the interview), primary school (six years completed), junior secondary (nine years completed), senior secondary (12 years completed), and higher education (more than 12 years). For the other countries, I use harmonized "educational attainment" variable from IPUMS-International. This variable does not necessarily reflect any particular country's education classification system but rather defines four roughly comparable educational categories: less than primary completed, primary completed, secondary completed, and university completed. Descriptive statistics on education levels of working age married men and women for all countries and time periods are presented in Table 3.

**Table 3. Education levels of working age married men and women**

| **Panel A: Brazil** | 1970 | | 2010 | |
|---|---|---|---|---|
| | Men | Women | Men | Women |
| Less than primary school | 91.7% | 93.9% | 41.4% | 33.8% |
| Primary school | 5.0% | 3.8% | 27.7% | 28.8% |
| Secondary school | 2.2% | 2.1% | 23.2% | 26.7% |
| University | 1.1% | 0.2% | 7.8% | 10.7% |
| Number of couples | | 2,433,218 | | 1,765,755 |
| **Panel B: Indonesia** | 1993 | | 2014 | |
| | Men | Women | Men | Women |
| Less than primary school | 37.5% | 45.5% | 12.5% | 11.3% |
| Primary school | 27.4% | 28.5% | 22.7% | 23.7% |
| Junior secondary school | 12.3% | 11.1% | 18.8% | 22.6% |
| Senior secondary school | 16.8% | 11.8% | 32.8% | 28.9% |
| Higher education | 6.0% | 3.1% | 13.3% | 13.6% |
| Number of couples | | 5,093 | | 8,442 |
| **Panel C: Mexico** | 1970 | | 2015 | |
| | Men | Women | Men | Women |
| Less than primary school | 77.3% | 80.5% | 22.1% | 21.2% |
| Primary school | 19.6% | 18.1% | 53.8% | 55.8% |
| Secondary school | 1.4% | 0.9% | 15.1% | 15.7% |
| University | 1.7% | 0.5% | 8.9% | 7.3% |
| Number of couples | | 39,639 | | 873,321 |
| **Panel D: South Africa** | 1996 | | 2011 | |
| | Men | Women | Men | Women |
| Less than primary school | 26.2% | 23.3% | 12.4% | 10.0% |
| Primary school | 46.4% | 49.7% | 38.9% | 41.0% |
| Secondary school | 22.9% | 23.8% | 36.6% | 37.3% |
| University | 4.6% | 3.2% | 12.1% | 11.7% |
| Number of couples | | 221,409 | | 251,042 |

The ideal measure of household labor income is the sum of the husband's and wife's gross monthly or annual income from wages and self-employment. Any income from government sources such as tax refunds, pensions, and other welfare benefits should be excluded because it tends to reduce labor market income inequality. Gross income data from wages and self-employment were available for Brazil, Indonesia, and Mexico. Income data for South Africa includes income from all sources (e.g., income from work, investments, welfare grants, remittances, rentals, etc.). This is not overly problematic, however, because almost three-quarters of all income in South Africa comes from wages, salaries, and self-employment (Stats SA 2011; Tregenna and Tsela 2012). Furthermore, given that the age restriction is 25–55 years old, pensions should not be an important source of income in the 2011 South African sample. To account for differences in household size and composition, household income is adjusted using the modified-OECD equivalence scale.[8] Household income distribution by household type for the later period (year 2) is presented in Table 4.

**Table 4. Relative household income by educational pairing**

| | Wife's education level | | | | |
|---|---|---|---|---|---|
| Husband's education level | LP | P | S | U | |
| | **Panel A: Brazil 2010 (N = 1,765,755 couples)** | | | | |
| LP | 0.361 | 0.481 | 0.717 | 1.310 | |
| P | 0.539 | 0.709 | 0.931 | 1.689 | |
| S | 0.742 | 0.896 | 1.280 | 2.322 | |
| U | 1.657 | 1.813 | 2.533 | 4.633 | |
| | **Panel B: Mexico 2015 (N = 873,321 couples)** | | | | |
| LP | 0.305 | 0.494 | 0.864 | 1.531 | |
| P | 0.459 | 0.731 | 1.166 | 1.987 | |
| S | 0.771 | 1.096 | 1.663 | 2.676 | |
| U | 1.640 | 1.721 | 2.534 | 3.892 | |
| | **Panel C: South Africa 2011 (N = 251,042 couples)** | | | | |
| LP | 0.146 | 0.147 | 0.365 | 1.747 | |
| P | 0.159 | 0.260 | 0.549 | 1.733 | |
| S | 0.431 | 0.495 | 1.236 | 2.624 | |
| U | 2.418 | 1.832 | 2.669 | 4.048 | |
| | **Panel D: Indonesia 2014 (N = 8,442 couples)** | | | | |
| | LP | P | JS | SS | HED |
| LP | 0.421 | 0.591 | 0.494 | 0.579 | 1.579 |
| P | 0.472 | 0.515 | 0.578 | 0.589 | 0.974 |
| JS | 0.511 | 0.702 | 0.834 | 0.902 | 0.936 |
| SS | 0.677 | 0.773 | 0.925 | 1.141 | 1.635 |
| HED | 1.826 | 1.287 | 1.196 | 1.688 | 2.746 |

*Notes:* Each cell denotes mean household income by educational pairing relative to mean income of all households in the sample. Household income is adjusted by the modified-OECD equivalence scale to account for differences in household size and composition.

[8] The modified-OECD equivalence scale distinguishes between adults and children by assigning a weight of 1 to the household head, 0.5 to each additional adult member aged 14 and older, and 0.3 to each child below age 14.

## 4. Findings

This section presents the main empirical findings, followed by robustness checks, comparisons with other relevant empirical studies, and a discussion of unexpected results explained through additional analysis of sorting trends by educational group.

### 4.1 Main findings

I find evidence of PAM at all levels of education in each country and sample year, consistent with Becker's prediction. The values on the main diagonal of the $I_{ij}$ tables in Tables 5–8 are all greater than 1, with some notably high values indicating significant levels of sorting. The degree of PAM is highest among the most educated individuals, which is especially evident in the 1970 Brazil sample where the $I_{ij}$ value reaches 53.19 for those with a university degree. Figure 1 complements the $I_{ij}$ tables by presenting evidence of overall PAM across all education levels in each country-year combination. The overall PAM is calculated as the weighted sum of the values along the main diagonal of the $I_{ij}$ table, using the marginal distribution of wives' education as weights.

The rise in educational attainment for both men and women between the two time periods in every country is evident from Table 3 in Section 3.2 and the changing marginal distributions of observed contingency tables shown in Panels A and B of Tables A.1–A.4 in Appendix A. Once these changes in the marginals are controlled for by standardizing the first-period tables to isolate changes in pure sorting from changes in educational attainment, I find that overall PAM has decreased in all countries (see Figure 2). The magnitude of the decrease varies across countries, ranging from 14% in Indonesia and South Africa to 23% in Mexico and 35% in Brazil.

The main results regarding the impact of educational PAM on household income inequality are summarized in Figure 3. In the first simulation, I find that household income inequality would be lower than observed if people matched randomly instead of assortatively. Although Figure 3 presents the simulation results only for the later period (year 2), this conclusion also applies to the earlier period (year 1) in each country. The counterfactual Gini coefficient under the random matching scenario is consistently lower than the observed Gini coefficient. For example, when random matching is imposed in the 2010 Brazil sample, the Gini coefficient decreases from 0.585 to 0.540, representing approximately an 8% decrease. This suggests that the observed level of household income inequality in the 2010 Brazil sample would have been lower by 8% if people matched randomly instead of assortatively.

In the second simulation, I examine what the level of household income inequality in the later period (year 2) would be if the degree of sorting were the same as in the earlier period (year 1). Given the hypothesis of a strong positive correlation between educational sorting and household income inequality, and the findings that overall PAM has decreased in all four countries, one would expect inequality to be significantly higher in year 2 if individuals sorted as they did in year 1. Contrary to expectations, I find a limited impact of changes in the degree of educational sorting on household income inequality. The counterfactual Gini coefficient barely differs from the observed Gini coefficient, with differences of only 1% in South Africa, 2% in Indonesia, and 3% in Brazil and Mexico. I further investigate why this might be the case in Section 4.4.

**Table 5.** $I_{ij}$ tables for Brazil (1970 and 2010)

| Husband's education level (*i*) | Wife's education level (*j*) | | | |
|---|---|---|---|---|
| | LP | P | S | U |
| | Panel A: 1970 (N = 2,433,218 couples) | | | |
| LP | 1.04 | 0.42 | 0.28 | 0.09 |
| P | 0.67 | 7.03 | 4.84 | 2.79 |
| S | 0.42 | 7.98 | 13.63 | 9.88 |
| U | 0.22 | 7.72 | 18.62 | 53.19 |
| | Panel B: 2010 (N = 1,765,755 couples) | | | |
| LP | 1.86 | 0.86 | 0.38 | 0.18 |
| P | 0.63 | 1.66 | 0.95 | 0.51 |
| S | 0.22 | 0.71 | 2.12 | 1.47 |
| U | 0.06 | 0.24 | 1.14 | 5.67 |

*Note:* Four educational categories are classified as less than primary school (LP), primary school (P), secondary school (S), and university (U).

**Table 6.** $I_{ij}$ tables for Indonesia (1993 and 2014)

| Husband's education level (*i*) | Wife's education level (*j*) | | | | |
|---|---|---|---|---|---|
| | LP | P | JS | SS | HED |
| | Panel A: 1993 (N = 5,093 couples) | | | | |
| LP | 1.74 | 0.62 | 0.24 | 0.05 | 0.03 |
| P | 0.92 | 1.61 | 0.80 | 0.26 | 0.14 |
| JS | 0.51 | 1.36 | 2.23 | 1.06 | 0.21 |
| SS | 0.20 | 0.84 | 2.09 | 3.26 | 1.73 |
| HED | 0.03 | 0.33 | 1.08 | 3.81 | 10.47 |
| | Panel B: 2014 (N = 8,442 couples) | | | | |
| LP | 4.15 | 1.42 | 0.62 | 0.18 | 0.05 |
| P | 1.31 | 1.95 | 1.11 | 0.42 | 0.14 |
| JS | 0.61 | 1.09 | 1.67 | 0.88 | 0.32 |
| SS | 0.20 | 0.50 | 0.97 | 1.75 | 0.99 |
| HED | 0.04 | 0.10 | 0.31 | 1.08 | 4.34 |

*Note:* Five educational categories are classified as less than primary school (LP), primary school (P), junior secondary (JS), senior secondary (SS), and higher education (HED).

**Table 7.** $I_{ij}$ tables for Mexico (1970 and 2015)

| Husband's education level (*i*) | Wife's education level (*j*) | | | |
|---|---|---|---|---|
| | LP | P | S | U |
| | Panel A: 1970 (N = 39,639 couples) | | | |
| LP | 1.15 | 0.37 | 0.32 | 0.59 |
| P | 0.53 | 3.04 | 1.99 | 0.99 |
| S | 0.22 | 3.88 | 10.63 | 4.45 |
| U | 0.17 | 3.70 | 12.48 | 16.67 |
| | Panel B: 2015 (N = 873,321 couples) | | | |
| LP | 2.62 | 0.75 | 0.14 | 0.03 |
| P | 0.73 | 1.27 | 0.73 | 0.28 |
| S | 0.16 | 0.80 | 2.62 | 1.51 |
| U | 0.04 | 0.30 | 2.03 | 6.89 |

*Note:* Four educational categories are classified as less than primary school (LP), primary school (P), secondary school (S), and university (U).

**Table 8.** $I_{ij}$ tables for South Africa (1996 and 2011)

| Husband's education level (*i*) | Wife's education level (*j*) | | | |
|---|---|---|---|---|
| | LP | P | S | U |
| | Panel A: 1996 (N = 221,409 couples) | | | |
| LP | 2.70 | 0.71 | 0.07 | 0.01 |
| P | 0.60 | 1.44 | 0.59 | 0.14 |
| S | 0.07 | 0.61 | 2.67 | 1.50 |
| U | 0.01 | 0.16 | 2.12 | 12.90 |
| | Panel B: 2011 (N = 251,042 couples) | | | |
| LP | 4.87 | 1.03 | 0.22 | 0.06 |
| P | 0.84 | 1.58 | 0.67 | 0.17 |
| S | 0.16 | 0.66 | 1.66 | 0.81 |
| U | 0.06 | 0.15 | 0.87 | 5.21 |

*Note:* Four educational categories are classified as less than primary school (LP), primary school (P), secondary school (S), and university (U).

**Figure 1.** Overall PAM in each country and sample year

*Notes:* Overall PAM is computed as the weighted sum of the values along the main diagonal of the $I_{ij}$ table (see Tables 5–8). The marginal distribution of wives' education is used as weights (see the last row ($f_{t+j}$) of the observed contingency tables in Panels A and B of Tables A.1–A.4 in Appendix A).

**Figure 2.** Time trends in overall PAM (year 1 standardized)

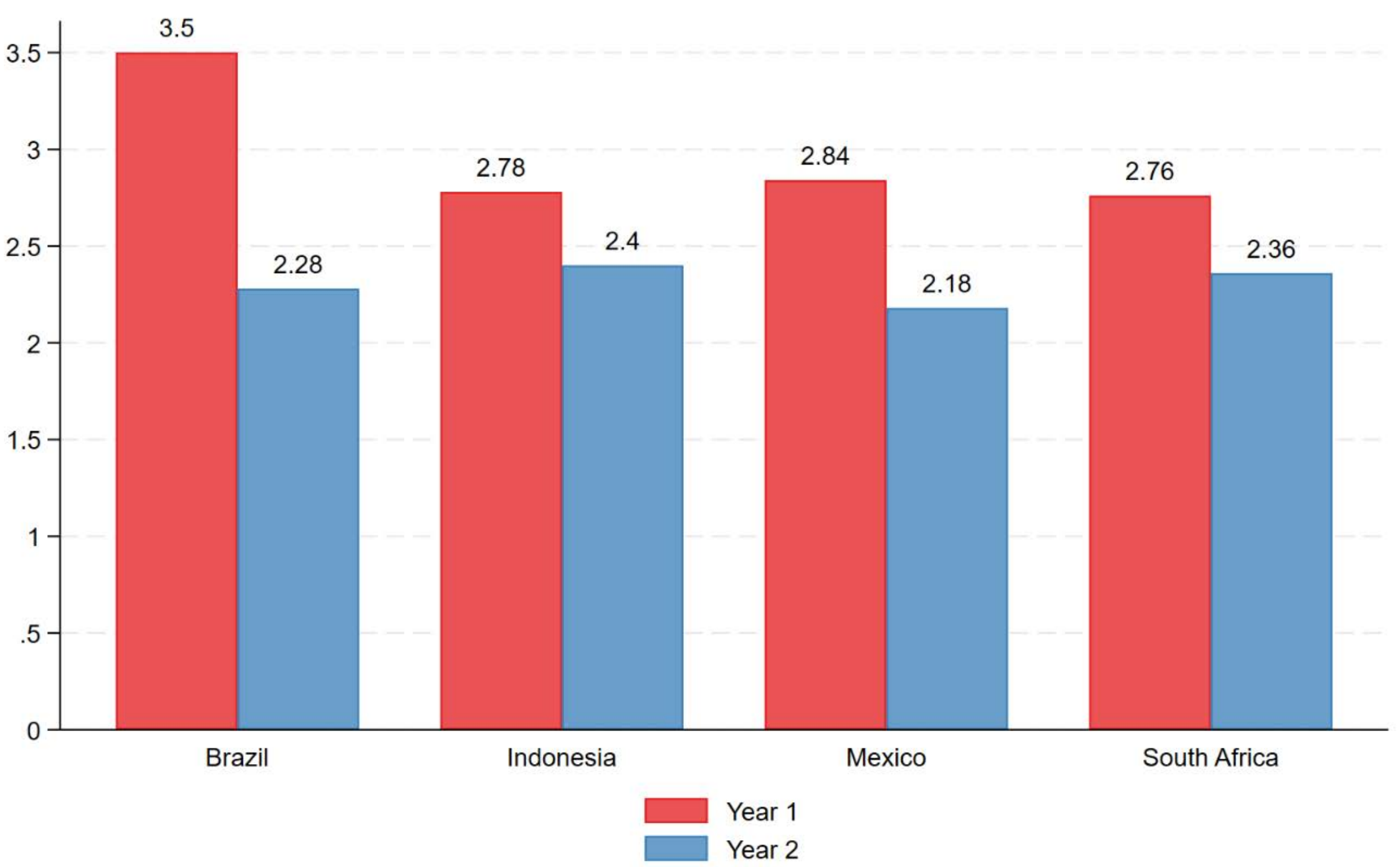

*Notes:* Overall PAM for Year 1 is calculated based on the first-period standardized contingency tables (see Panel C of Tables A.1–A.4 in Appendix A). For Year 2, overall PAM is calculated using the second-period observed tables (see Panel B of Tables A.1–A.4 in Appendix A).

**Figure 3.** Observed and simulated levels of household income inequality

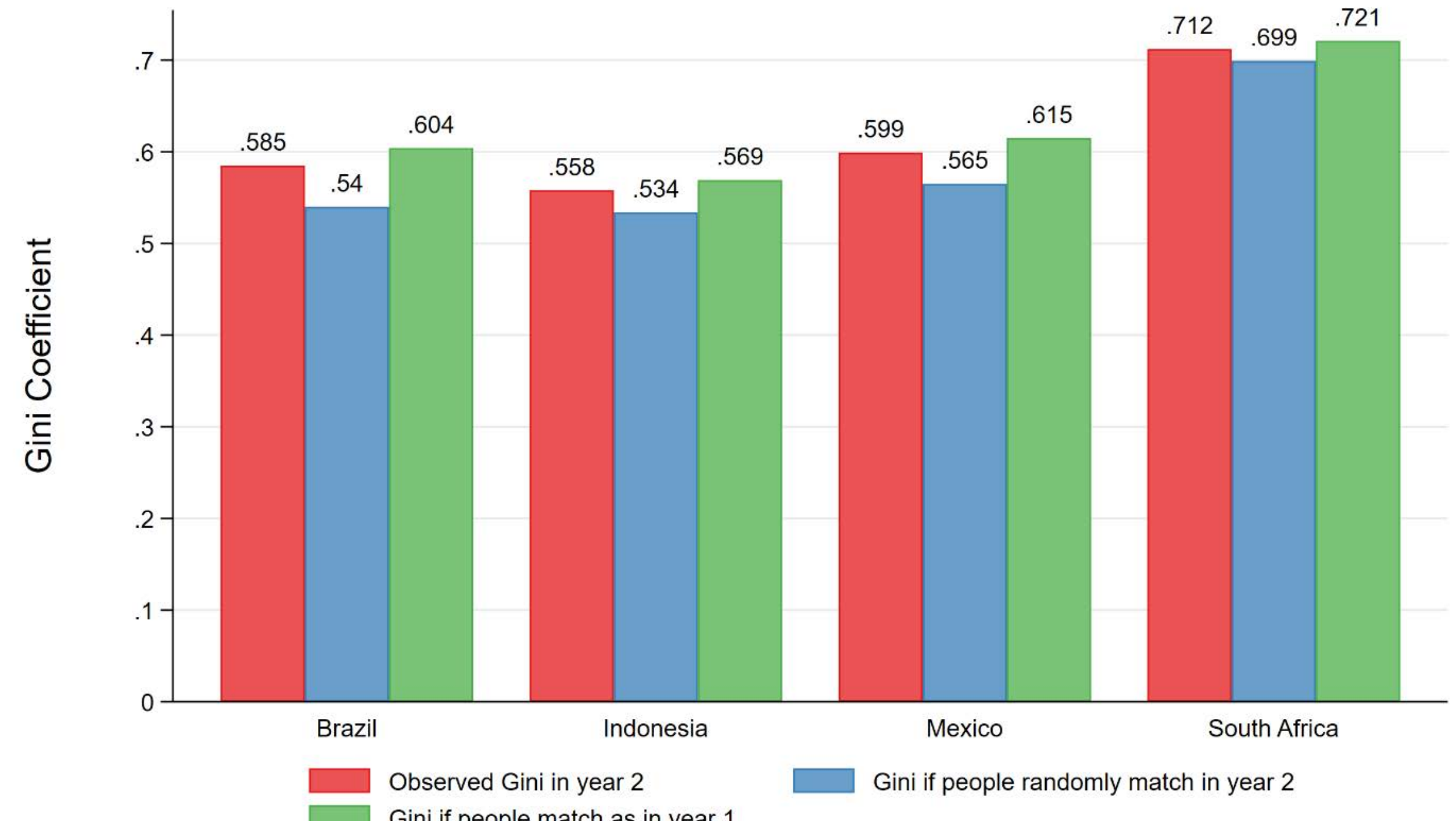

### 4.2 Robustness checks

In time trend analysis of overall PAM, the decision on which observed contingency table to standardize to control for changes in the marginals is somewhat arbitrary. Similarly, the choice between using the marginal distribution of wives' or husbands' education, or any convex combination of the two, as weights for calculating overall PAM is also arbitrary. To address these concerns, I conduct several robustness checks to examine the sensitivity of the trend direction shown in Figure 2, with the results reported in Appendix A.

In the first two robustness checks, I standardize the first-period observed contingency table, following the approach illustrated in Figure 2, but employ two different weighting strategies to calculate overall PAM. Figure A.1 in Appendix A presents the results of the first robustness check, where I use the marginal distribution of husbands' education as weights for calculating overall PAM. The magnitude of the decrease across countries is almost identical to those depicted in Figure 2, ranging from 13% in Indonesia and South Africa to 23% in Mexico and 34% in Brazil. The results of the second robustness check, where I use the average of the husbands' and wives' marginals as weights, are shown in Figure A.2 in Appendix A. Once again, the results are almost identical to those in Figure 2, indicating that the trend direction and magnitude depicted in Figure 2 are robust to the choice of different weights for calculating overall PAM.

In the subsequent three robustness checks, I standardize the second-period observed contingency table and then calculate overall PAM using three different weighting strategies.[9] As shown in Figures A.3–A.5 in Appendix A, there is a consistent downward trend in overall PAM across countries. While the magnitude of the decrease varies somewhat from that in Figure 2, the ranking of countries remains similar, with Brazil and Mexico exhibiting the most substantial decreases. These results underscore the robustness of the trend depicted in Figure 2, not only across different weighting schemes but also across different standardization choices.

[9] The standardized second-period tables are reported in Panel D of Tables A.1–A.4 in Appendix A.

### 4.3 Comparison with two methodologically related studies

As detailed in Kujundzic (2024), the literature on educational sorting lacks consensus regarding sorting trends. The only general conclusion one can make is that there is a large variation across studies about trend direction and magnitude, even among studies using the same dataset. This variability arises mainly because of the margin-dependency problem of sorting measures used in different studies, resulting in biased estimates of sorting trends that confound changes in the marginal distribution of education with changes in pure sorting.

This ongoing lack of consensus complicates the comparison of this study's findings with other empirical studies of educational assortative mating and income inequality. The most relevant comparisons are two important papers by Greenwood et al. (2014, 2015) and Eika et al. (2019), which are the focus of this section. There are two main reasons for this.

First, Eika et al. (2019) also use the interest factor to measure the degree of PAM for each educational group and overall PAM across all educational groups. The interest factor is a margin-dependent measure of association, meaning it is sensitive to changes in the marginals. Therefore, it is important to control for changes in the marginals to identify pure sorting trends and their potential impact on household income inequality. Eika and coauthors claim to control for changes in the marginals, although they do not explicitly discuss their methodology (Eika et al. 2019, p. 2811). Additionally, the specific weighting strategy used to calculate overall PAM is unclear. Nonetheless, their findings on overall PAM trends should be comparable to this study's findings, as trend direction is robust to the choice of different weights if changes in the marginals are controlled for. Second, Greenwood et al. (2014) paper is also relevant because this study closely follows their methodology to simulate the impact of educational PAM on household income inequality.

Greenwood et al. (2014) utilize IPUMS census data from 1960 to 2005 to analyze educational marriage patterns and household income inequality in the US. They use three different measures of overall educational PAM: the regression coefficient (γ), Kendall's tau (τ) rank correlation, and the ratio of the sum of diagonal elements of the observed and random matching contingency tables (δ).[10] All of these measures have a margin-dependency problem, leading to inconsistent trends in overall PAM depending on the measure used.

Kendall's τ exhibits non-monotonic trends, indicating that overall educational PAM increased from 1960 to 1980, decreased until 2000, and then increased again from 2000 to 2005. While both γ and δ suggest an increase in overall PAM from 1960 to 1980, their trends diverge after 1980: δ remains relatively stable from 1980 to 2000 before increasing again, whereas γ continues to rise from 1980 to 2000 and then stabilizes.

Due to these inconsistent results, along with potential computational issues not explicitly addressed, the authors revised their initial conclusions in a 2015 corrigendum. Their original assertion about changes in the degree of sorting over time having a substantial impact on household income inequality was deemed incorrect. The corrected results[11] from the two counterfactual experiments reported in the corrigendum align with the findings of this study: while household income inequality would be lower if people matched randomly instead of assortatively, changes in the degree of sorting over time have a limited impact on household income inequality.

[10] According to Almar and Schulz (2023), δ is mathematically equivalent to the weighted sum of the values along the main diagonal of the $I_{ij}$ table, without using an explicit weighting scheme.

[11] The authors do not discuss how they corrected the original 2014 analysis to arrive at the new results.

Eika et al. (2019) also explore long-run educational sorting trends and household income inequality in the US. Employing data from the US decennial censuses for the 1940–1962 period and the March Current Population Survey (CPS) for the 1962–2013 period, they use the interest factor (I) to measure the degree of PAM for each educational group and overall PAM across all groups. Their findings reveal evidence of PAM at all levels of education throughout the entire period, which is in line with the results of this study. However, their findings on trends in overall PAM in the US differ from those of this study for four middle-income countries. Specifically, the authors note a steady increase in overall PAM in the US from 1940 to the mid-1980s, followed by relatively little change. In contrast, this study identifies a consistent downward trend in overall PAM from the 1970s to the 2010s in Brazil and Mexico and from the 1990s to the 2010s in Indonesia and South Africa.

To quantify the potential impact of educational PAM on household income inequality, Eika and coauthors employ a decomposition method proposed by DiNardo et al. (1996) using the CPS data. Their findings align with those of Greenwood et al. (2015) and the results of this study: household income inequality would be lower if people matched randomly instead of assortatively, but changes in the degree of sorting over time have minimal impact on household income inequality. Furthermore, the authors provide additional evidence demonstrating that these results for the US extend to Norway for the 1967–2013 period.

### 4.4 Trends in PAM by educational group

Given these findings, a natural question to ask is why changes in overall educational PAM have such a small impact on household income inequality. Additional analysis reveals that this counterintuitive result is due to differing educational group dynamics. As shown in Figure 4, the degree of PAM by educational group varies in magnitude and direction. In all four countries, highly educated individuals sort among themselves the most, whereas the least educated sort the least. However, although the level of sorting is highest among the most educated group, it has been decreasing over time. Meanwhile, sorting among the least educated group has been increasing in all countries except Brazil. This indicates that the inequality-decreasing impact of a decrease in sorting among the highly educated is almost completely offset by the inequality-increasing impact of an increase in sorting among the least educated group. Consequently, the overall PAM has a small impact on household income inequality. Interestingly, Eika et al. (2019) reach the same conclusion for the US (1962–2013) and Norway (1967–2013).

In the case of Brazil, all educational groups show a decrease in sorting, but with varying magnitudes. The 35% decrease in overall PAM is mostly driven by a strong decrease in sorting among those with a university degree, while sorting levels remain relatively stable for the other three groups. However, the size of the university group is much smaller than the combined size of the other three groups. This means that the inequality-decreasing impact of reduced sorting among the university group is not sufficiently large to induce significant changes in the household income distribution. As a result, a 35% decrease in overall PAM leads to only a 3% reduction in household income inequality.

**Figure 4.** Time trends in PAM by education level

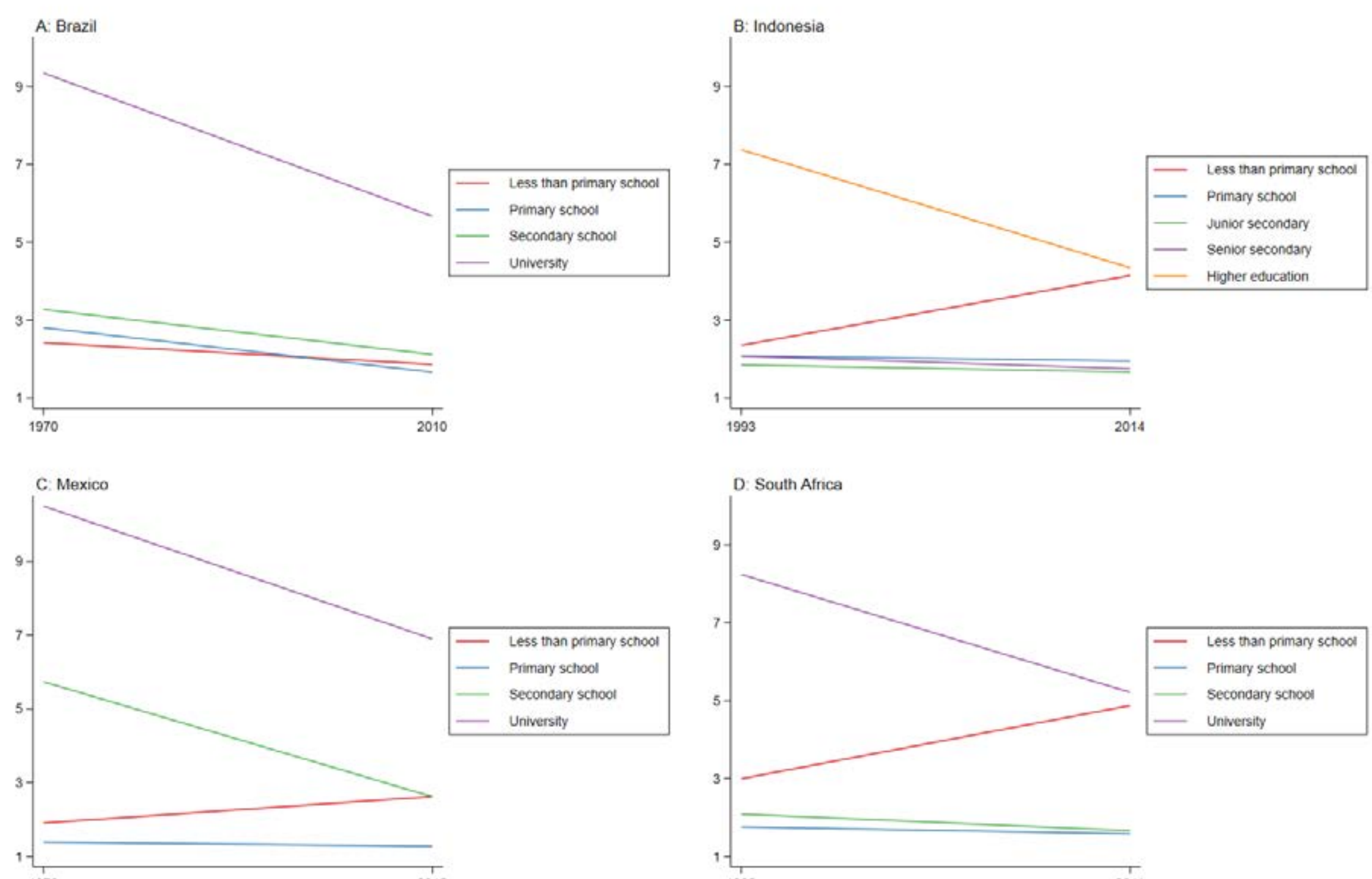

*Notes:* Observed first-period contingency tables are standardized to match the marginals of the observed second-period tables using the quadratic programming (QP) method described in Appendix B. The standardized first-period tables are reported in Panel C of Tables A.1–A.4 in Appendix A, while the observed second-period tables are reported in Panel B of the same tables. The measure of PAM for each educational group is calculated using the formula $I_{ij} = f_{t_{ij}} / f_{t_{i+}} f_{t_{+j}}$ (for $i = j$). This calculation is performed using the standardized first-period tables for the earlier period and the observed second-period tables for the later period.

## 5. Summary and concluding remarks

This paper provides new empirical evidence of educational PAM and its implications for household income inequality from four emerging economies: Brazil (1970–2010), Indonesia (1993–2014), Mexico (1970–2015), and South Africa (1996–2011). Middle-income countries tend to exhibit higher levels of household income inequality than high-income countries and have experienced substantial educational expansion over the past few decades. Furthermore, these four countries together are home to roughly 40–50 million people living in extreme poverty, representing about 6–7% of the 727 million people worldwide who lived on less than $1.90 a day in 2015 (UN World Social Report 2020). Taken together, these facts underscore the importance of examining the potential impact of educational sorting on household income inequality in middle-income countries – not only from a distributional perspective but also from a poverty-reduction perspective, a dimension that much of the existing literature has largely overlooked in its focus on high-income country contexts.

The first major conclusion of this study supports Becker's theory of sorting, demonstrating that individuals tend to form internally homogenous marriages based on education level. This finding aligns with most other empirical studies of educational assortative mating across various contexts (e.g., Schwartz and Mare 2005; Blossfeld 2009; Greenwood et al. 2014; Boertien and Permanyer 2019; Eika et al. 2019).

The second major conclusion is that highly educated individuals tend to sort among themselves the most, while the least educated sort the least. Although the highest level of sorting is among the most educated individuals, this trend has been decreasing over time. Conversely, sorting among the least educated has been increasing in all countries except Brazil. These sorting trends among different groups are consistent with findings by Eika et al. (2019) for a set of high-income countries: Denmark (1980–2013), Germany (1984–2013), Norway (1967–2013), the United Kingdom (1979–2013), and the US (1940–2013).

The third major conclusion is that overall sorting across all educational groups has decreased in all countries once changes in educational attainment are controlled for by standardizing one of the observed contingency tables. Failure to control for changes in educational attainment results in biased estimates of sorting trends, as it becomes impossible to distinguish between changes in pure sorting and changes in educational attainment. This is perhaps best illustrated in the case of Brazil. By examining Figure 1 in Section 4.1, we might erroneously conclude that the overall level of educational sorting in Brazil has increased from 1.65 in 1970 to 2.28 in 2010. This apparent increase in sorting is driven by both changes in pure sorting and changes in educational attainment. Once the changes in educational attainment are controlled for by standardizing the first-period contingency table, we see that sorting has, in fact, decreased from 3.5 in 1970 to 2.28 in 2010, as shown in Figure 2 in Section 4.1. This suggests that educational expansion is largely responsible for the results observed in Figure 1.

Building on these insights, I conduct two thought experiments to test Becker's hypothesis that educational sorting can exacerbate income inequality between households. The results indicate that while household income inequality would be lower if people matched randomly instead of assortatively, changes in the degree of sorting over time have a limited impact on household income inequality. This conclusion is in line with findings by Greenwood et al. (2015) for the US and Eika et al. (2019) for the US and Norway.

Further analysis reveals that this counterintuitive result stems from differing educational group dynamics. In all four countries, highly educated individuals are sorting less among themselves, whereas the least educated are increasingly sorting among themselves in all countries except Brazil. This suggests that the inequality-decreasing impact of reduced sorting among the highly educated is almost completely offset by the inequality-increasing impact of increased sorting among the least educated group. Consequently, changes in the overall degree of sorting across all educational groups have had a minimal impact on household income inequality.

While it is certainly reassuring that concerns about educational assortative mating having a potentially large effect on income disparities between households appear to be unwarranted, these findings suggest another concerning narrative. Marginalization processes are occurring at low levels of educational distribution. The least educated are being left behind, facing limited labor market opportunities and diminished chances of achieving upward socioeconomic mobility through marriage to more educated partners.

The primary advantage of the methodology employed in this paper is its flexibility, as it allows for the non-parametric identification of pure sorting trends without having to rely on any specific theoretical model to describe the sorting process. My main consideration in deciding on the identification strategy was to avoid the rather restrictive assumption regarding the role of search frictions in the matching process. Most theoretical models of sorting can be classified into two general categories based on this key assumption: frictionless matching models and search models. Frictionless matching models assume a frictionless marketplace where there are no costs associated with the search for partners or with obtaining additional information about their characteristics. In contrast, search models incorporate search frictions, acknowledging that the process of finding a long-term partner can be costly and time-consuming, especially in the absence of complete information about prospective partners' characteristics. Although search models are more realistic, many empirical applications are based on the frictionless matching framework due to the greater complexity and difficulty in estimating search models empirically (e.g., Chiappori et al. 2017; Ciscato and Weber 2020; Dupuy and Weber 2022).

That said, the flexibility of not having to rely on any specific theoretical framework to identify pure sorting trends comes at the cost of not being able to distinguish between the potential mechanisms driving the observed sorting patterns in the data. The key mechanisms include individual preferences, search frictions, and resource allocation within marriages. This limitation impacts the interpretation of the empirical findings in this study. Specifically, the findings about the tendency of individuals to form internally homogenous marriages based on education level, as well as changes over time, should not be interpreted strictly as statements about marital preferences.

With this in mind, a suggestion for future research is to leverage additional data sources beyond traditional census data that offer insights into preferences or help minimize search frictions. One approach could be to apply the methodology used in this paper to survey data that includes detailed information about marital preferences. Another approach is to utilize online dating data. According to Hitsch et al. (2010), online dating data could be useful for this exercise due to the minimal search frictions and the availability of information about the choice sets faced by users (e.g., information about a potential partner's age, education, and job profile) and the choices they make from these choice sets. However, Hitsch and coauthors do not observe whether two users who met online went on an actual date or eventually got married. Instead, they can only infer the likelihood of a date based on information exchanged online. Therefore, to pursue the proposed line of inquiry, one would need to obtain online dating data that includes information about marriage formation after meeting online, if available.

## References

Acemoglu, Daron, and David Autor. 2011. "Skills, Tasks and Technologies: Implications for Employment and Earnings." In *Handbook of Labor Economics*, edited by Orley Ashenfelter and David Card, 4:1043–1171. Elsevier.

Acemoglu, Daron, and Jonas Loebbing. 2024. *Automation and Polarization*. Working Paper, MIT Department of Economics. Available at: https://economics.mit.edu/people/faculty/daron-acemoglu/working-papers.

Acemoglu, Daron, and Pascual Restrepo. 2022. "Tasks, Automation, and the Rise in US Wage Inequality." *Econometrica* 90 (5): 1973–2016.

Acemoglu, Daron, and Pascual Restrepo. 2024. *Automation and Rent Dissipation: Implications for Wages, Inequality, and Productivity*. Working Paper, MIT Department of Economics. Available at: https://economics.mit.edu/people/faculty/daron-acemoglu/working-papers.

Almar, Frederik, and Bastian Schulz. 2023. *Optimal Weights for Marital Sorting Measures.* IZA Discussion Paper no. 16368. Bonn: Institute for the Study of Labor. Available at: https://docs.iza.org/dp16368.pdf

Balestra, Carlotta, Ana Llena-Nozal, Fabrice Murtin, Elena Tosetto, and Benoît Arnaud. 2018. *Inequalities in Emerging Economies: Informing the Policy Dialogue on Inclusive Growth.* OECD Statistics Working Papers, 2018/13. OECD Publishing, Paris.

Becker, Gary S. 1973. "A theory of marriage: Part I." *Journal of Political Economy* 81 (4): 813–846.

Becker, Gary S. 1974. "A theory of marriage: Part II." *Journal of Political Economy* 82 (2, Part 2): S11–S26.

Berg, Andrew, Jonathan D. Ostry, Charalambos G. Tsangarides, and Yorbol Yakhshilikov. 2018. "Redistribution, inequality, and growth: new evidence." *Journal of Economic Growth* 23: 259–305.

Blossfeld, Hans-Peter. 2009. "Educational assortative marriage in comparative perspective." *Annual Review of Sociology* 35: 513–530.

Breen, Richard, and Leire Salazar. 2011. "Educational assortative mating and earnings inequality in the United States." *American Journal of Sociology* 117 (3): 808–843.

Boertien, Diederik, and Iñaki Permanyer. 2019. "Educational assortative mating as a determinant of changing household income inequality: A 21-country study." *European Sociological Review* 35 (4): 522–537.

Bourguignon, François. 2015. *The Globalization of Inequality*. Princeton University Press.

Chiappori, Pierre-André, Bernard Salanié, and Yoram Weiss. 2017. "Partner Choice, Investment in Children, and the Marital College Premium." *American Economic Review* 107 (8): 2109–2167.

Ciscato, Edoardo, and Simon Weber. 2020. "The Role of Evolving Marital Preferences in Growing Income Inequality." *Journal of Population Economics* 33 (1): 307–347.

Daly, Mary C., and Robert G. Valletta. 2006. "Inequality and Poverty in United States: The Effects of Rising Dispersion of Men's Earnings and Changing Family Behaviour." *Economica* 73 (289): 75–98.

De Hauw, Yolien, André Grow, and Jan Van Bavel. 2017. "The Reversed Gender Gap in Education and Assortative Mating in Europe." *European Journal of Population* 33 (4): 445–474.

DiNardo, John, Nicole M. Fortin, and Thomas Lemieux. 1996. "Labor Market Institutions and the Distribution of Wages, 1973-1992: A Semiparametric Approach." *Econometrica* 64 (5): 1001–1044.

Dupuy, Arnaud, and Simon Weber. 2022. "Marriage Market Counterfactuals Using Matching Models." *Economica* 89 (353): 29–43.

Eika, Lasse, Magne Mogstad, and Basit Zafar. 2019. "Educational Assortative Mating and Household Income Inequality." *Journal of Political Economy* 127 (6): 2795–2835.

Fernandez, Raquel, Nezih Guner, and John Knowles. 2005. "Love and Money: A Theoretical and Empirical Analysis of Household Sorting and Inequality." *The Quarterly Journal of Economics* 120 (1): 273–344.

Goldberg, Pinelopi Koujianou, and Nina Pavcnik. 2007. "Distributional Effects of Globalization in Developing Countries." *Journal of Economic Literature* 45 (1): 39–82.

Greenwood, Jeremy, Nezih Guner, Georgi Kocharkov, and Cezar Santos. 2014. "Marry Your Like: Assortative Mating and Income Inequality." *American Economic Review* 104 (5): 348–353.

Greenwood, Jeremy, Nezih Guner, Georgi Kocharkov, and Cezar Santos. 2015. Corrigendum to "Marry Your Like: Assortative Mating and Income Inequality." Available at: http://www.jeremygreenwood.net/papers/ggksPandPcorrigendum.pdf.

Herre, Bastian, Veronika Samborska, Esteban Ortiz-Ospina, and Max Roser. 2020. "Marriages and Divorces." Published online at OurWorldinData.org. Retrieved from: 'https://ourworldindata.org/marriages-and-divorces' [Online Resource]

Hitsch, Günter J., Alι Hortaçsu, and Dan Ariely. 2010. "Matching and Sorting in Online Dating." *American Economic Review* 100 (1): 130–163.

Hoehn-Velasco, Lauren, and Jacob Penglase. 2023. "Changes in Assortative Matching and Educational Inequality: Evidence from Marriage and Birth Records in Mexico." *Journal of Demographic Economics* 89 (4): 587–607.

Hryshko, Dmytro, Chinhui Juhn, and Kristin McCue. 2017. "Trends in earnings inequality and earnings instability among US couples: How important is assortative matching?" *Labour Economics* 48: 168–182.

Hyslop, Dean R., and David C. Maré. 2005. "Understanding New Zealand's changing income distribution, 1983–1998: A semi-parametric analysis." *Economica* 72 (287): 469–495.

Kujundzic, Ana. 2024. "Measuring Trends in Assortative Mating." Chapter 2 of PhD thesis *Understanding Income and Labor Market Inequalities: Methods and Applications*. Available at: https://doi.org/10.18174/676190

Minnesota Population Center. 2020. Integrated Public Use Microdata Series, International: Version 7.3. Minneapolis, MN: IPUMS. Available at: https://doi.org/10.18128/D020.V7.3.

Pencavel, John. 1998. “Assortative mating by schooling and the work behavior of wives and husbands.” *The American Economic Review* 88 (2): 326–329.

Ritchie, Hannah, Veronika Samborska, Natasha Ahuja, Esteban Ortiz-Ospina, and Max Roser. 2023. “Global Education.” Published online at OurWorldinData.org. Retrieved from: 'https://ourworldindata.org/global-education' [Online Resource]

Rodrik, Dani. 1999. “Where did all the growth go? External shocks, social conflict, and growth collapses.” *Journal of Economic Growth* 4 (4): 385–412.

Samudra, Rizky, and I Dewa Gede Karma Wisana. 2015. “Love Between Us: Educational Assortative Mating and Expenditure Inequality in Indonesia.” Universitas Indonesia. Available here.

Sandel, Michael J. 2020. *The Tyranny of Merit: What's Become of the Common Good?* Penguin UK.

Schwartz, Christine R., and Robert D. Mare. 2005. “Trends in Educational Assortative Marriage from 1940 to 2003.” *Demography* 42 (4): 621–646.

Stats SA. 2011. *Statistics South Africa*. General Household Survey.

Strauss, John, and Firman Witoelar. 2021. *Indonesia Family Life Survey*. In *Encyclopedia of Gerontology and Population Aging*, edited by D. Gu and M.E. Dupre. Springer, Cham. Available at: https://doi.org/10.1007/978-3-030-22009-9_339.

Tregenna, Fiona, and Mfanafuthi Tsela. 2012. “Inequality in South Africa: The distribution of income, expenditure and earnings.” *Development Southern Africa* 29 (1): 35–61.

United Nations. Department of Economic and Social Affairs. 2020. *World social report 2020: Inequality in a rapidly changing world*. UN. Available here.

Zhang, Yuanting, and Jennifer Van Hook. 2009. “Marital dissolution among interracial couples.” *Journal of Marriage and Family* 71 (1): 95–107.

# Appendices

## Appendix A. Additional Tables and Figures

**Table A.1.** Observed and standardized contingency tables for Brazil (1970 and 2010)

| Husband's education level ($f_{t_i}$) | Wife's education level ($f_{t_j}$) | | | | Total ($f_{t_{i+}}$) |
|---|---|---|---|---|---|
| | LP | P | S | U | |
| Panel A: observed 1970 table | | | | | |
| LP | 0.897 | 0.015 | 0.005 | 0.000 | 0.917 |
| P | 0.032 | 0.013 | 0.005 | 0.000 | 0.050 |
| S | 0.009 | 0.007 | 0.006 | 0.001 | 0.022 |
| U | 0.002 | 0.003 | 0.004 | 0.001 | 0.011 |
| Total ($f_{t_{+j}}$) | 0.939 | 0.038 | 0.021 | 0.002 | $f_{t_{++}}$ = 1.000 |
| Panel B: observed 2010 table | | | | | |
| LP | 0.261 | 0.103 | 0.042 | 0.008 | 0.414 |
| P | 0.059 | 0.132 | 0.070 | 0.015 | 0.277 |
| S | 0.017 | 0.047 | 0.131 | 0.037 | 0.232 |
| U | 0.002 | 0.005 | 0.024 | 0.047 | 0.078 |
| Total ($f_{t_{+j}}$) | 0.338 | 0.288 | 0.267 | 0.107 | $f_{t_{++}}$ = 1.000 |
| Panel C: standardized 1970 table | | | | | |
| LP | 0.338 | 0.064 | 0.011 | 0.000 | 0.414 |
| P | 0.000 | 0.223 | 0.053 | 0.000 | 0.277 |
| S | 0.000 | 0.000 | 0.203 | 0.029 | 0.232 |
| U | 0.000 | 0.000 | 0.000 | 0.078 | 0.078 |
| Total ($f_{t_{+j}}$) | 0.338 | 0.288 | 0.267 | 0.107 | $f_{t_{++}}$ = 1.000 |
| Panel D: standardized 2010 table | | | | | |
| LP | 0.863 | 0.034 | 0.018 | 0.002 | 0.917 |
| P | 0.046 | 0.003 | 0.001 | 0.000 | 0.050 |
| S | 0.021 | 0.001 | 0.001 | 0.000 | 0.022 |
| U | 0.010 | 0.000 | 0.000 | 0.000 | 0.011 |
| Total ($f_{t_{+j}}$) | 0.939 | 0.038 | 0.021 | 0.002 | $f_{t_{++}}$ = 1.000 |

*Notes:* Each cell in Panels A and B represents the observed share (or relative frequency) of household type $t_{ij}$, where husbands have education level *i* and wives have education level *j*. Panel C displays the observed 1970 contingency table that has been standardized so that the row-total marginal ($f_{t_{i+}}$) and the column-total marginal ($f_{t_{+j}}$) match those of the observed 2010 table in Panel B. Panel D shows the observed 2010 table that has been standardized to have the same marginals as the observed 1970 table in Panel A. The standardization is done using the quadratic programming (QP) procedure described in Kujundzic (2024).

**Table A.2.** Observed and standardized contingency tables for Indonesia (1993 and 2014)

| Husband's education level ($f_{t_i}$) | Wife's education level ($f_{t_j}$) | | | | | Total ($f_{t_{i+}}$) |
|---|---|---|---|---|---|---|
| | LP | P | JS | SS | HED | |
| | Panel A: observed 1993 table | | | | | |
| LP | 0.296 | 0.066 | 0.010 | 0.002 | 0.000 | 0.375 |
| P | 0.114 | 0.126 | 0.024 | 0.008 | 0.001 | 0.274 |
| JS | 0.028 | 0.048 | 0.030 | 0.015 | 0.001 | 0.123 |
| SS | 0.015 | 0.040 | 0.039 | 0.065 | 0.009 | 0.168 |
| HED | 0.001 | 0.006 | 0.007 | 0.027 | 0.020 | 0.060 |
| Total ($f_{t_{+j}}$) | 0.455 | 0.285 | 0.111 | 0.118 | 0.031 | $f_{t_{++}}$ = 1.000 |
| | Panel B: observed 2014 table | | | | | |
| LP | 0.058 | 0.042 | 0.017 | 0.006 | 0.001 | 0.125 |
| P | 0.034 | 0.104 | 0.057 | 0.027 | 0.004 | 0.227 |
| JS | 0.013 | 0.048 | 0.071 | 0.048 | 0.008 | 0.188 |
| SS | 0.007 | 0.038 | 0.072 | 0.166 | 0.044 | 0.328 |
| HED | 0.001 | 0.003 | 0.009 | 0.041 | 0.078 | 0.133 |
| Total ($f_{t_{+j}}$) | 0.113 | 0.237 | 0.226 | 0.289 | 0.136 | $f_{t_{++}}$ = 1.000 |
| | Panel C: standardized 1993 table | | | | | |
| LP | 0.033 | 0.041 | 0.022 | 0.028 | 0.000 | 0.125 |
| P | 0.040 | 0.112 | 0.045 | 0.029 | 0.000 | 0.227 |
| JS | 0.018 | 0.057 | 0.079 | 0.035 | 0.000 | 0.188 |
| SS | 0.022 | 0.027 | 0.080 | 0.197 | 0.003 | 0.328 |
| HED | 0.000 | 0.000 | 0.000 | 0.000 | 0.133 | 0.133 |
| Total ($f_{t_{+j}}$) | 0.113 | 0.237 | 0.226 | 0.289 | 0.135 | $f_{t_{++}}$ = 1.000 |
| | Panel D: standardized 2014 table | | | | | |
| LP | 0.288 | 0.041 | 0.020 | 0.018 | 0.008 | 0.375 |
| P | 0.062 | 0.141 | 0.038 | 0.027 | 0.006 | 0.274 |
| JS | 0.035 | 0.044 | 0.025 | 0.017 | 0.002 | 0.123 |
| SS | 0.046 | 0.047 | 0.024 | 0.045 | 0.007 | 0.168 |
| HED | 0.025 | 0.013 | 0.004 | 0.010 | 0.008 | 0.061 |
| Total ($f_{t_{+j}}$) | 0.455 | 0.285 | 0.111 | 0.117 | 0.031 | $f_{t_{++}}$ = 1.000 |

*Notes:* Panel C shows the observed 1993 contingency table that has been standardized to have the same marginals as the observed 2014 table in Panel B. Panel D shows the observed 2014 table that has been standardized to have the same marginals as the observed 1993 table in Panel A.

**Table A.3.** Observed and standardized contingency tables for Mexico (1970 and 2015)

| Husband's education level ($f_{t_i}$) | Wife's education level ($f_{t_j}$) | | | | Total ($f_{t_{i+}}$) |
|---|---|---|---|---|---|
| | LP | P | S | U | |
| | Panel A: observed 1970 table | | | | |
| LP | 0.717 | 0.052 | 0.002 | 0.002 | 0.773 |
| P | 0.083 | 0.108 | 0.004 | 0.001 | 0.196 |
| S | 0.002 | 0.010 | 0.001 | 0.000 | 0.014 |
| U | 0.002 | 0.012 | 0.002 | 0.001 | 0.017 |
| Total ($f_{t_{+j}}$) | 0.805 | 0.181 | 0.009 | 0.005 | $f_{t_{++}}$ = 1.000 |
| | Panel B: observed 2015 table | | | | |
| LP | 0.123 | 0.093 | 0.005 | 0.001 | 0.221 |
| P | 0.084 | 0.382 | 0.062 | 0.011 | 0.538 |
| S | 0.005 | 0.067 | 0.062 | 0.017 | 0.151 |
| U | 0.001 | 0.015 | 0.028 | 0.045 | 0.089 |
| Total ($f_{t_{+j}}$) | 0.212 | 0.558 | 0.157 | 0.073 | $f_{t_{++}}$ = 1.000 |
| | Panel C: standardized 1970 table | | | | |
| LP | 0.089 | 0.132 | 0.000 | 0.000 | 0.221 |
| P | 0.123 | 0.415 | 0.000 | 0.000 | 0.538 |
| S | 0.000 | 0.011 | 0.136 | 0.004 | 0.151 |
| U | 0.000 | 0.000 | 0.021 | 0.069 | 0.089 |
| Total ($f_{t_{+j}}$) | 0.212 | 0.558 | 0.157 | 0.073 | $f_{t_{++}}$ = 1.000 |
| | Panel D: standardized 2015 table | | | | |
| LP | 0.637 | 0.126 | 0.007 | 0.003 | 0.773 |
| P | 0.143 | 0.051 | 0.002 | 0.000 | 0.196 |
| S | 0.011 | 0.002 | 0.000 | 0.000 | 0.014 |
| U | 0.015 | 0.002 | 0.000 | 0.001 | 0.017 |
| Total ($f_{t_{+j}}$) | 0.805 | 0.181 | 0.009 | 0.005 | $f_{t_{++}}$ = 1.000 |

*Notes:* Panel C shows the observed 1970 contingency table that has been standardized to have the same marginals as the observed 2015 table in Panel B. Panel D shows the observed 2015 table that has been standardized to have the same marginals as the observed 1970 table in Panel A.

**Table A.4.** Observed and standardized contingency tables for South Africa (1996 and 2011)

| Husband's education level ($f_{t_i}$) | Wife's education level ($f_{t_j}$) | | | | Total ($f_{t_{i+}}$) |
|---|---|---|---|---|---|
| | LP | P | S | U | |
| | Panel A: observed 1996 table | | | | |
| LP | 0.165 | 0.093 | 0.004 | 0.000 | 0.262 |
| P | 0.065 | 0.332 | 0.065 | 0.002 | 0.464 |
| S | 0.004 | 0.069 | 0.145 | 0.011 | 0.229 |
| U | 0.000 | 0.004 | 0.023 | 0.019 | 0.046 |
| Total ($f_{t_{+j}}$) | 0.233 | 0.497 | 0.238 | 0.032 | $f_{t_{++}}$ = 1.000 |
| | Panel B: observed 2011 table | | | | |
| LP | 0.060 | 0.053 | 0.010 | 0.001 | 0.124 |
| P | 0.033 | 0.251 | 0.097 | 0.008 | 0.389 |
| S | 0.006 | 0.099 | 0.227 | 0.034 | 0.366 |
| U | 0.001 | 0.007 | 0.039 | 0.074 | 0.121 |
| Total ($f_{t_{+j}}$) | 0.100 | 0.410 | 0.373 | 0.117 | $f_{t_{++}}$ = 1.000 |
| | Panel C: standardized 1996 table | | | | |
| LP | 0.037 | 0.066 | 0.021 | 0.000 | 0.124 |
| P | 0.045 | 0.281 | 0.063 | 0.000 | 0.389 |
| S | 0.018 | 0.064 | 0.285 | 0.000 | 0.366 |
| U | 0.000 | 0.000 | 0.005 | 0.117 | 0.121 |
| Total ($f_{t_{+j}}$) | 0.100 | 0.410 | 0.373 | 0.117 | $f_{t_{++}}$ = 1.000 |
| | Panel D: standardized 2011 table | | | | |
| LP | 0.230 | 0.026 | 0.003 | 0.003 | 0.262 |
| P | 0.000 | 0.339 | 0.112 | 0.012 | 0.464 |
| S | 0.000 | 0.111 | 0.109 | 0.008 | 0.229 |
| U | 0.003 | 0.021 | 0.014 | 0.008 | 0.046 |
| Total ($f_{t_{+j}}$) | 0.233 | 0.497 | 0.238 | 0.032 | $f_{t_{++}}$ = 1.000 |

*Notes:* Panel C shows the observed 1996 contingency table that has been standardized to have the same marginals as the observed 2011 table in Panel B. Panel D shows the observed 2011 table that has been standardized to have the same marginals as the observed 1996 table in Panel A.

**Figure A.1.** Time trends in overall PAM (year 1 standardized, weights are the male marginal)

*Notes:* Overall PAM for Year 1 is calculated based on the first-period standardized contingency tables shown in Panel C of Tables A1–A4. For Year 2, overall PAM is calculated using the second-period observed tables shown in Panel B of Tables A1–A4. The marginal distribution of husbands' education is used as weights (see the last column ($f_{t_{i+}}$) in Panels B and C of Tables A1–A4).

**Figure A.2.** Time trends in overall PAM (year 1 standardized, weights are the average of male and female marginals)

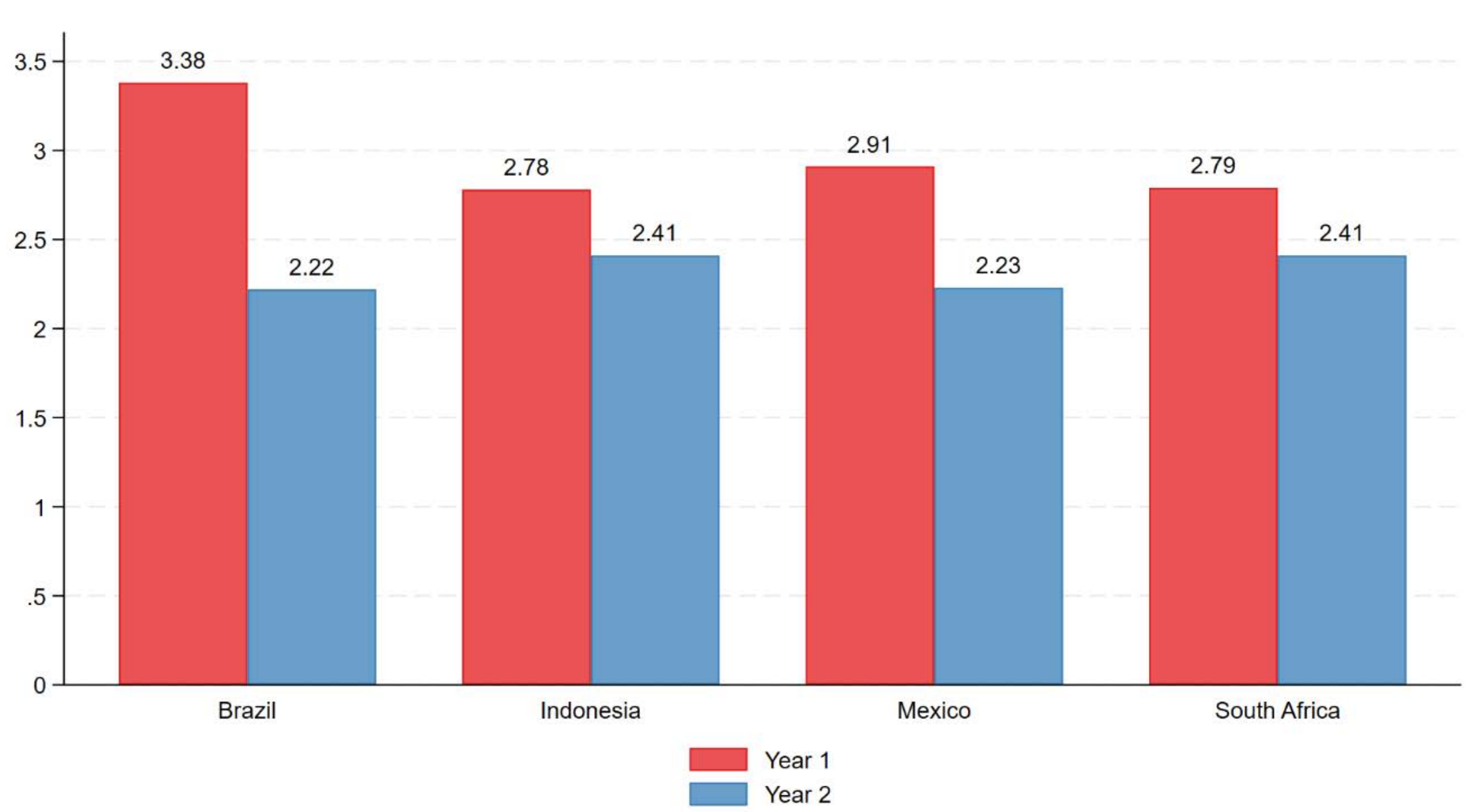

*Note*: Overall PAM is calculated as in Figure A1, except that the average of the husbands' and wives' marginals are used as weights.

**Figure A.3.** Time trends in overall PAM (year 2 standardized, weights are the female marginal)

*Notes:* Overall PAM for Year 1 is calculated based on the first-period observed contingency tables shown in Panel A of Tables A1–A4. For Year 2, overall PAM is calculated using the second-period standardized tables shown in Panel D of Tables A1–A4. The marginal distribution of wives' education is used as weights (see the last row ($f_{t_{+j}}$) in Panels A and D of Tables A1–A4).

**Figure A.4.** Time trends in overall PAM (year 2 standardized, weights are the male marginal)

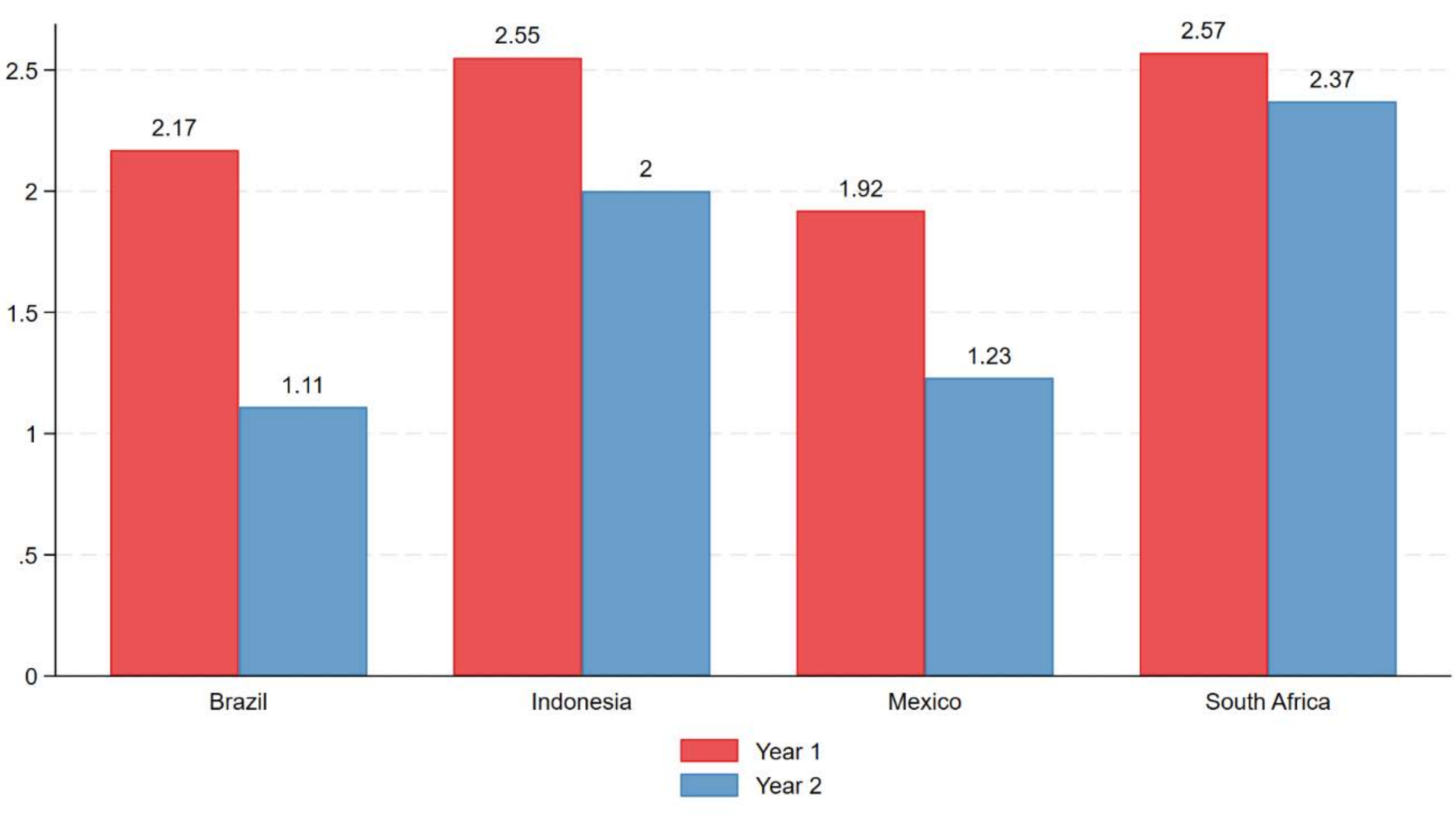

*Note*: Overall PAM is calculated as in Figure A3, except that the husbands' marginal is used as weights (see the last column ($f_{t_{i+}}$) in Panels A and D of Tables A1–A4).

**Figure A.5.** Time trends in overall PAM (year 2 standardized, weights are the average of male and female marginals)

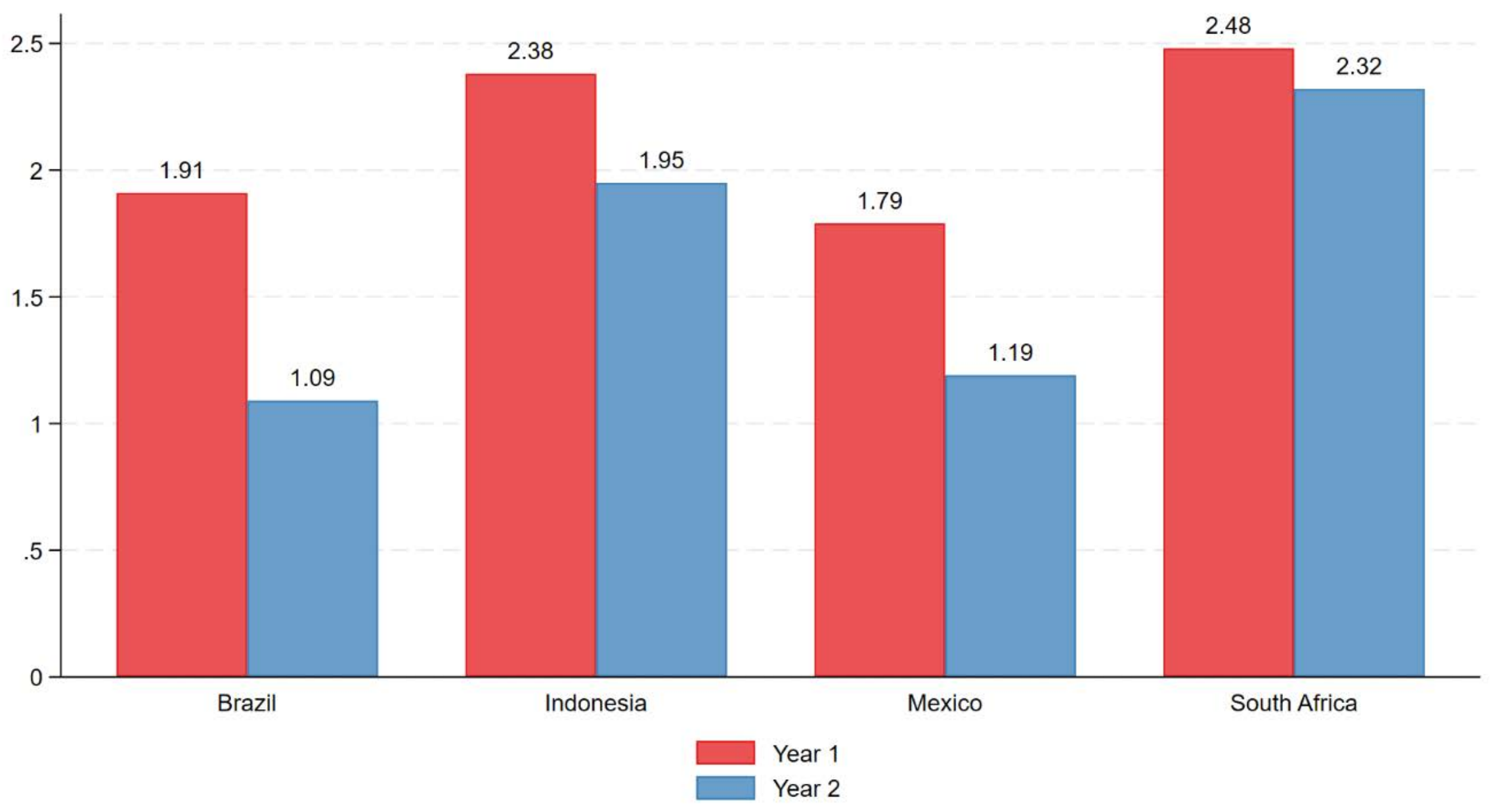

*Note*: Overall PAM is calculated as in Figure A3, except that the average of the husbands' and wives' marginals are used as weights.

## Appendix B. Quadratic Programming (QP) Standardization Method

The following explanation of the QP standardization method is taken from Kujundzic (2024).

Suppose we have two contingency tables, $M$ and $N$, that we want to compare. Both tables have the same dimensions $(I \times J)$ but different marginals. The idea behind the QP method is to standardize table $M$ so that it has the same marginals as table $N$, while preserving the core pattern of association of table $M$, as measured by local interest factors ($I_{ij}$). This standardization problem is formulated as a quadratic programming problem, where the quadratic objective function is the distance between the observed and counterfactual $I_{ij}$ values of table $M$, and the linear constraints are the marginals of table $N$.

The inner cells of an $(I \times J)$ contingency table can be expressed as relative frequencies instead of frequency counts:

$$f_{ij}^{r} = \frac{f_{ij}}{f_{++}} \tag{B.1}$$

where $f_{ij}^{r}$ is the relative frequency of the $(i, j)$ cell, $f_{ij}$ is the frequency count of the $(i, j)$ cell, and $f_{++}$ is the total number of observations.

The local interest factor for the $(i, j)$ cell of an $(I \times J)$ contingency table is defined as:

$$I_{ij} = \frac{f_{ij}^{r}}{f_{i+}^{r} \times f_{+j}^{r}} \tag{B.2}$$

where $f_{i+}^{r} = \sum_j f_{ij}^{r}$ is the row $i$ total, and $f_{+j}^{r} = \sum_i f_{ij}^{r}$ is the column $j$ total.

From Equation (B.2), it follows that:

$$\sum_i \left(I_{ij} \times f_{i+}^{r}\right) = 1, \qquad for\ all\ j \tag{B.3}$$

and

$$\sum_j \left(I_{ij} \times f_{+j}^{r}\right) = 1, \qquad for\ all\ i \tag{B.4}$$

The objective is to find a counterfactual $I_{ij}$ (denoted by $\tilde{I}_{ij}$) that is as close as possible to the observed $I_{ij}$ of table *M*, in order to preserve the core pattern of association of table *M*, while ensuring that the marginal conditions of table *N* are satisfied. If "closeness" is defined as a quadratic function of the distance between the counterfactual and observed $I_{ij}$, then the standardization problem can be formulated as a quadratic programming problem of the following form:

$$\min_{\{\tilde{I}_{ij}\}} \sum_{ij} \left(\tilde{I}_{ij} - I_{ij}^{M}\right)^2 \tag{B.5}$$

subject to

$$\sum_{i} \left(\tilde{I}_{ij} \times f_{i+}^{r\,N}\right) = 1, \qquad for\ all\ j \tag{B.6}$$

$$\sum_{j} \left(\tilde{I}_{ij} \times f_{+j}^{r\,N}\right) = 1, \qquad for\ all\ i \tag{B.7}$$

$$\tilde{I}_{ij} \geq 0 \tag{B.8}$$

where Equations (B.6) and (B.7) represent the row-sum and column-sum marginal conditions of table *N* that must be satisfied, and Equation (B.8) specifies non-negativity restrictions.

Under standard conditions, the QP method always converges to a solution that satisfies the marginal constraints, provided the target marginals are internally consistent and define a non-empty feasible set (i.e., a set of non-negative cell values that satisfy the specified marginals). Moreover, since the objective function is strictly convex and the constraints are linear, the optimization problem yields a unique and well-defined standardized contingency table.

**Working Papers of Eesti Pank 2026**

No 1
Erwan Gautier, Cristina Conflitti, Daniel Enderle, Ludmila Fadejeva, Alex Grimaud, Eduardo Gutiérrez, Valentin Jouvanceau, Jan-Oliver Menz, Alari Paulus, Pavlos Petroulas, Pau Roldan-Blanco, Elisabeth Wieland. Consumer Price Stickiness in the Euro Area During an Inflation Surge